\documentclass{article}

\usepackage{microtype}
\usepackage{multirow}
\usepackage{graphicx}
\usepackage{arydshln}
\usepackage{xcolor}
\usepackage{subcaption}
\usepackage{booktabs} % for professional tables

\usepackage{hyperref}

\definecolor{mygreen}{RGB}{0,160,0}
\newcommand{\g}[1]{\textcolor{mygreen}{#1}}

\definecolor{myred}{RGB}{220,0,0}
\newcommand{\red}[1]{\textcolor{myred}{#1}}

\definecolor{myred_dark}{RGB}{130,0,0}
\newcommand{\reddark}[1]{\textcolor{myred_dark}{#1}}

\newcommand{\algcomment}[1]{\hfill {\color{blue} $\triangleright$ \emph{\small{#1}}}}
\usepackage[accepted]{icml2024}

\usepackage{amsmath}
\usepackage{amssymb}
\usepackage{mathtools}
\usepackage{amsthm}

\usepackage[capitalize,noabbrev]{cleveref}

\theoremstyle{plain}

\theoremstyle{definition}

\theoremstyle{remark}

\usepackage[textsize=tiny]{todonotes}

\icmltitlerunning{Explaining the Model, Protecting Your Data}

\begin{document}

\twocolumn[
\icmltitle{\textit{Explaining the Model, Protecting Your Data}: Revealing and Mitigating the Data Privacy Risks of Post-Hoc Model Explanations via Membership Inference}

% It is OKAY to include author information, even for blind
% submissions: the style file will automatically remove it for you
% unless you've provided the [accepted] option to the icml2024
% package.

% List of affiliations: The first argument should be a (short)
% identifier you will use later to specify author affiliations
% Academic affiliations should list Department, University, City, Region, Country
% Industry affiliations should list Company, City, Region, Country

% You can specify symbols, otherwise they are numbered in order.
% Ideally, you should not use this facility. Affiliations will be numbered
% in order of appearance and this is the preferred way.
\icmlsetsymbol{equal}{*}

\begin{icmlauthorlist}
\icmlauthor{Catherine Huang}{yyy}
\icmlauthor{Martin Pawelczyk}{yyy}
\icmlauthor{Himabindu Lakkaraju}{yyy}
\end{icmlauthorlist}

\icmlaffiliation{yyy}{Harvard University}
\icmlcorrespondingauthor{Catherine Huang}{catherinehuang@college.harvard.edu}
% \icmlcorrespondingauthor{Martin Pawelczyk}{martin.pawelczyk.1@gmail.com}
% \icmlcorrespondingauthor{Hima Lakkaraju}{hlakkaraju@hbs.edu}

% You may provide any keywords that you
% find helpful for describing your paper; these are used to populate
% the "keywords" metadata in the PDF but will not be shown in the document
\icmlkeywords{Explainability, Interpretability, Post-Hoc Explanations, Privacy, Data Privacy, Differential Privacy, Foundation Models, Deep Learning, Trustworthy ML, Membership Inference Attacks, Adversarial ML}

\vskip 0.3in
]

% this must go after the closing bracket ] following \twocolumn[ ...

% This command actually creates the footnote in the first column
% listing the affiliations and the copyright notice.
% The command takes one argument, which is text to display at the start of the footnote.
% The \icmlEqualContribution command is standard text for equal contribution.
% Remove it (just {}) if you do not need this facility.

%\printAffiliationsAndNotice{}  % leave blank if no need to mention equal contribution
\printAffiliationsAndNotice{\icmlEqualContribution} % otherwise use the standard text.

\begin{abstract}
Predictive machine learning models are becoming increasingly deployed in high-stakes contexts involving sensitive personal data; in these contexts, there is a trade-off between model \textit{explainability} and data \textit{privacy}. In this work, we push the boundaries of this trade-off: with a focus on foundation models for image classification fine-tuning, we reveal unforeseen privacy risks of post-hoc model explanations and subsequently offer mitigation strategies for such risks. First, we construct VAR-LRT and L1/L2-LRT, two new membership inference attacks based on feature attribution explanations that are significantly more successful than existing explanation-leveraging attacks, particularly in the low false-positive rate regime that allows an adversary to identify specific training set members with confidence. Second, we find empirically that optimized differentially private fine-tuning substantially diminishes the success of the aforementioned attacks, while maintaining high model accuracy. We carry out a systematic empirical investigation of our 2 new attacks with 5 vision transformer architectures, 5 benchmark datasets, 4 state-of-the-art post-hoc explanation methods, and 4 privacy strength settings.
\end{abstract}

% % This analysis fills a gap in literature---there is no prior work that thoroughly quantifies the relationship between differential privacy and the subsequent privacy risks of post-hoc explanations in a deep learning setting. 

\section{Introduction}
\label{section:introduction}
Predictive machine learning (ML) models are becoming increasingly deployed in high-stakes contexts such as medical diagnoses and loan approvals. Since these models rely on sensitive personal data, regulatory principles that enforce safe and trustworthy model training and usage have become increasingly important. One key regulatory principle is the \textit{Right to Privacy}, which aims to protect against training data leakage \cite{right_to_privacy}. The right to privacy is a limit to model \textit{explainability}, which is itself another important pillar of trustworthy ML. Given the inherent complexity of high-stakes models, explanations are increasingly necessary in offering users information about how models make decisions with respect to data. It is common to use \textit{post-hoc} explanations that explain the behavior of a trained model on a \textit{specific} data example; \textit{feature attributions} are a broad and commonly studied sub-class of post-hoc explanations \cite{shap, lime, shrikumar}. 

One widely used standard to empirically verify whether a model obeys privacy is membership inference attacks (MIAs) \cite{original_mi_paper}, which predict if a data example was used to train a target model. Successful MIAs are a violation of privacy---if an adversary knows that a patient's medical record was used to train a model that predicts the optimal treatment for a particular disease, the adversary can correctly conclude that this patient has the disease.

There is limited work on the susceptibility of model explanations to membership inference, let alone work on the privacy risks of model explanations altogether. Our work reveals unforeseen data privacy violations of post-hoc feature attribution explanations through addressing the following question: Can we devise new membership inference attacks on feature attribution explanations that 1) have higher success than existing explanation-leveraging attacks, and 2) allow an adversary to confidently identify specific members of the training set in the ``low false-positive rate regime"?

% data privacy violation as a significant risk in the widespread adoption of 

Furthermore, despite the literature on the privacy risks of model explanations, there has been little work on actually \textit{mitigating} these risks through privacy-preserving methods. We hypothesize that the mathematical framework \textit{differential privacy (DP)} \cite{dwork} can be used during training to mitigate privacy violations of explanations, and we address this under-explored problem through the following question: Does differential privacy defend against an adversary's ability to leverage model explanations to infer sensitive training data membership information, while preserving model accuracy?

\section{Related Work} \label{section:related_work}
Prior work shows that explanations do risk leaking sensitive training data information via membership inference: \citet{featureattack} and \citet{recourseattack} show that backpropagation-based explanations and algorithmic recourse, respectively, can leak training set membership information. However, this existing work is limited: Shokri et al.'s attacks are evaluated using average-case metrics that do not characterize whether the attack can confidently identify any specific members of the training set. Pawelczyk et al.'s work highlights that an adversary can accurately identify specific training set members with high confidence but focuses only on counterfactual explanations of binary classification models, thus not addressing a broader class of feature attribution explanations on more complex deep classification models. Moreover, both works involve training low-dimensional real-world datasets; neither addresses the privacy risks of explanations coming from models trained or fine-tuned on complex datasets.

\citet{liu} develop a attack that trains a shadow model to infer membership based on how the target model behaves differently when the input is perturbed according to its feature attribution explanation. This recent work addresses the limitations of the aforementioned works by systematically demonstrating strong attack success, particularly at identifying specific training data members, in a deep learning setting.

In response to work on attacks against deep learning models, privacy-preserving deep learning has risen in importance as a research area. Albeit limited, there exists emergent work on differentially private (DP) computation of post-hoc explanations. \citet{huang2023accurate} propose and evaluate two methods to generate DP counterfactual explanations in logistic regression models. Even though Huang et al. find that DP reduces what an adversary can infer about training set membership, this work is limited specifically to counterfactual explanations of logistic regression models. They do not address privacy-preserving explanations in deep learning settings (in which privacy is a greater concern) or the broader class of post-hoc feature attribution explanations.

\textbf{Where Our Work Stands} \quad Our work extends that of Shokri et al. and Pawelczyk et al. by developing successful MIAs, leveraging feature attribution explanations on deep image classification models, that can confidently identify specific training set members at low FPR. Our attacks differ from Liu et al.'s in terms of adversarial information access: Liu et al.'s attack requires access to target model losses and entire attribution vectors, but our attack assumes no access to losses and utilizes only one-dimensional summaries of attributions. We extend Huang et al.'s work by offering a unified evaluation, which current literature lacks, of the impact of DP model training on feature attribution-based MIA performance in deep learning settings. Furthermore, our attacks are on foundation models and require no training from scratch, not even of the \textit{shadow models} used in the attacks---we conduct membership inference from \textit{fine-tuning} alone, a less computationally expensive process.

\section{Preliminaries} \label{section:preliminaries}

Let $D_{train} = \{\mathcal{X}, \mathcal{Y}\} = \{(\mathbf{x_i}, y_i)\}_{i=1}^N$ be a training dataset drawn from some underlying distribution $\mathbb{D}$. Let $f_{\theta}$ be the model parameterized by $\theta$, $\mathbf{x} \in \mathbb{R}^d$ be an input feature vector, and $y \in [k]$ be an output label. $\mathcal{X} \in \mathbb{R}^{N \times d}$ denotes the feature set, and $\mathcal{Y} \in [k]^N$ denotes the labels over $\mathcal{X}$.

\textbf{The Case for Foundation Models} \quad In this work, we evaluate pre-trained vision transformers on image classification fine-tuning tasks---membership inference, in our case, seeks to infer whether an example was used to \textit{fine-tune} the model. Appendix \ref{appendix:vision_transformers} discusses the choice of studying foundation models and specifically the vision transformer architecture. 

\textbf{Post-Hoc Feature Attribution Explanations} \quad A \textbf{post-hoc explanation} function $\varphi$ takes as input a trained model $f_{\theta}$ and a point of interest $\mathbf{x} \in \mathbb{R}^d$. A \textbf{feature attribution} post-hoc explanation $\varphi(\mathbf{x})$ is a $k$-dimensional vector whose $i$-th coordinate, $\varphi_i(\mathbf{x})$, reflects the extent to which the $i$-th feature influences the prediction the model outputs for $\mathbf{x}$. We study the following feature attribution methods: Input $*$ Gradient (IXG), Saliency Maps (SL), Integrated Gradients (IG), and (a gradient-based approximation to) SHAP (GS). We describe each of these methods in Appendix \ref{appendix:post_hoc_explanations}.

\textbf{Membership Inference Attacks} \quad Suppose an adversary possesses a set of data examples. The goal of a \textbf{membership inference attack (MIA)} is for an adversary to create a function that predicts whether each data example belongs to the training set of $f_{\theta}$. MIAs are predominantly loss-based, testing if the loss of the model for each example is below some threshold; we note that loss-based attacks require adversarial access to true labels. Traditionally, MIAs are evaluated using average-case metrics such as the receiver operating characteristic (ROC) curve---which plots attack true positive rate (TPR) against false positive rate (FPR)---and the area under that curve (AUC).

\textbf{Likelihood Ratio Attacks and the Low-FPR Regime} \quad
\citet{carlini} propose a re-formulation of the MIA problem to focus not on average-case performance but rather on the ``low FPR regime." If an MIA has high TPR at low FPR, that means it can \textit{confidently} identify the training set membership of a few observations in a sensitive dataset. Attack success at low FPR is a greater privacy violation than an attack that only \textit{unreliably} achieves high aggregate success rate. Carlini et al. also initiated the practice of reporting \textit{log-scaled} ROC curves, rather than linearly scaled curves, to make visible TPRs at very low FPRs.

Carlini et al. additionally propose the \textbf{Likelihood Ratio Attack (LiRA)} that is significantly more successful, in particular at low FPRs, than prior MIAs \cite{original_mi_paper, yeom, jayaraman, song_and_mittal, ye, watson2022importancedifficultycalibrationmembership, sablayrolles2019whiteboxvsblackboxbayes, long}. In LiRA, the adversary trains shadow models on datasets with and without target example $(\mathbf{x}, y)$. Let $\mathbb{Q}_{in}(\mathbf{x}, y) = \{f \gets \mathcal{T}(D_{attack} \cup \{(\mathbf{x},y)\} \ | \ D_{attack} \gets \mathbb{D}\}\}$ represent the distribution of models trained on datasets containing $(\mathbf{x}, y)$. Likewise, we have $\mathbb{Q}_{out}(\mathbf{x}, y) = \{f \gets \mathcal{T}(D_{attack} \setminus \{(\mathbf{x},y)\} \ | \ D_{attack} \gets \mathbb{D}\}\}$. The adversary estimates the likelihood ratio $\hat{\Lambda}(f_{\theta}; (\mathbf{x}, y)) \approx \frac{p(f_{\theta} | \mathbb{Q}_{in}(\mathbf{x}, y))}{p(f_{\theta} | \mathbb{Q}_{out}(\mathbf{x}, y))}$ and then thresholds on $\hat{\Lambda}$:
$\text{Membership}_{\text{LiRA}, \tau}(\mathbf{x}, y) = 
  \text{True} \text{ if $\hat{\Lambda} \ge \tau$}, \ 
  \text{False}  \text{ otherwise}$, where $\tau$ is a threshold that maximizes TPR at a given FPR. 

\textbf{Explanation-Based MIAs} \quad \citet{featureattack} propose an explanation-based attack that directly thresholds on the explanation variance. Example $\mathbf{x}$ is predicted to be a member of the training set iff $\mathrm{Var}(\varphi(\mathbf{x})) \le \tau,$ where $\tau$ is an optimal threshold we assume that the adversary has access to. We elaborate on the intuition behind using explanation variance in MIAs in Section \ref{section:methods}. 

\textbf{Differential Privacy} \quad DP is a mathematically provable definition of privacy that provides a quantifiable metric of an algorithm's privacy loss, providing a computational method whose output is random enough to obscure any single participant's presence in the training data \cite{dwork}. DP mechanisms have an $\varepsilon$ parameter that quantifies privacy strength (the lower the $\varepsilon$, the stronger the privacy strength) and a $\delta$ parameter that indicates the probability of privacy failure. In Appendix \ref{appendix:more_on_dp}, we explain the formal definition of DP and the optimized DP-stochastic gradient descent method by \citet{bu2023automatic} that we use in our experiments.

\section{Our Membership Inference Attack Methods on Model Explanations} \label{section:methods}
We present our new MIAs, drawn from Carlini et al.'s LiRA framework, that leverage the \textit{variances, L1 norms, and L2 norms} of each example's feature attribution. We name these attacks VAR-LRT, L1-LRT, and L2-LRT, respectively. These black-box attacks assume that for every example, the adversary has access to the model's prediction on that example and a post-hoc explanation; unlike in loss-based attacks, adversarial access to true labels is not required.

\textbf{Attack on Explanation Variances (VAR-LRT)} \quad \citet{featureattack} thresholding attack on explanation variance follows the intuition that gradient descent pushes training set points further from the decision boundary, and non-training points are on average closer to the decision boundary. (This intuition is also leveraged in other adversarial ML work \cite{choquettechoo2021labelonly, yu2019new}.) The act of leveraging explanation variance is motivated by this idea---that for points closer to the decision boundary, changing a feature affects the prediction itself more strongly, which leads to higher explanation variance. If a point is farther from the decision boundary, that means the model is more certain about the point's prediction, and the model's behavior on the point is unlikely to change if we slightly perturb the point. Shokri's attack method directly thresholds on explanation variance in inferring training set membership of each example; \textbf{we use the attack's intuition but enhance the attack's design.} Our first attack, VAR-LRT, computes and thresholds on likelihood ratios of explanation variances. Algorithm \ref{alg:variancesattack} shows VAR-LRT in detail.

\begin{algorithm}[htb!]
 \begin{algorithmic}[1] 
  \REQUIRE \text{model} $f_{\theta}$, \text{example} $(\mathbf{x}, y) \in \mathbb{R}^d$, \text{explanation vector} $\varphi(f, (\mathbf{x}, y)) \in \mathbb{R}^d$, \text{data distribution} $\mathbb{D}$, \text{number of shadow model iterations} $N_S$
  \STATE $\text{variances}_{\text{in}} = \{\}$, $\text{variances}_{\text{out}} = \{\}$
  \FOR{$N_S$ times}
    \STATE $D_{\text{attack}} \gets^\$ \mathbb{D}$ \algcomment{sample a shadow dataset}
    \STATE $f_{\text{in}} \gets \mathcal{T}(D_{\text{attack}} \cup \{(\mathbf{x},y)\})$ \algcomment{train IN model with $(\mathbf{x}, y)$ in training set}
    \STATE $\varphi_{\text{in}} \gets \varphi(f_{\text{in}}, (\mathbf{x}, y))$
    \algcomment{generate post-hoc explanation of $f_{\text{in}}$'s behavior on $(\mathbf{x}, y)$}
    \STATE $\bar{\varphi}_{\text{in}} \gets \frac{1}{d} \sum_{i=1}^{d} \varphi_{\text{in}, i}$
    \STATE $\text{variances}_{\text{in}} \gets \text{variances}_{\text{in}} \cup \{ \frac{1}{d}\sum_{i=1}^{d} (\varphi_{\text{in}, i} - \bar{\varphi}_{\text{in}})^2 \}$
    \algcomment{record sample variance of $\varphi_{\text{in}}$}
    \STATE $f_{\text{out}} \gets \mathcal{T}(D_{\text{attack}} {\setminus \{(\mathbf{x},y)\}})$ \algcomment{train OUT model}
    \STATE $\varphi_{\text{out}} \gets \varphi(f_{\text{out}}, (\mathbf{x}, y))$
    \algcomment{generate post-hoc explanation of $f_{\text{out}}$'s behavior on $(\mathbf{x}, y)$}
    \STATE $\bar{\varphi}_{\text{out}} \gets \frac{1}{d} \sum_{i=1}^{d} \varphi_{\text{out}, i}$
    \STATE $\text{variances}_{\text{out}} \gets \text{variances}_{\text{out}} \cup \{ \frac{1}{d}\sum_{i=1}^{d} (\varphi_{\text{out}, i} - \bar{\varphi}_{\text{out}})^2 \}$
     \algcomment{record sample variance of $\varphi_{\text{out}}$}
  \ENDFOR
  \STATE $\hat{\mu}_{\text{in}} \gets \texttt{mean}(\text{variances}_{\text{in}})$, $\hat{\mu}_{\text{out}} \gets \texttt{mean}(\text{variances}_{\text{out}})$
  %{1 \over N}\sum_{i=1}^N \text{loss}_{\text{in}}$
  \STATE $\hat{\sigma}_{\text{in}}^2 \gets \texttt{var}(\text{variances}_{\text{in}})$,  $\hat{\sigma}_{\text{out}}^2 \gets \texttt{var}(\text{variances}_{\text{out}})$
  %{1 \over N}\sum_{i=1}^N \text{loss}_{\text{in}}$
  \STATE $\varphi_{\text{obs}} \gets \varphi(f_{\theta}, (\mathbf{x}, y))$, \ $\bar{\varphi}_{\text{obs}} \gets \sum_{i=1}^{d} \varphi_{\text{obs}, i}$
  \STATE $\text{variance}_{\text{obs}} = \frac{1}{d}\sum_{i=1}^{d} (\varphi_{\text{obs}, i} - \bar{\varphi}_{\text{obs}})^2$ \algcomment{query model}
  \vspace{0.5em}
  \STATE \textbf{return} $\displaystyle \hat{\Lambda} = \frac{p(\text{variance}_{\text{obs}}\ \mid\ \mathcal{N}(\hat{\mu}_{\text{in}}, \hat{\sigma}^2_{\text{in}}))}
    { p(\text{variance}_{\text{obs}}\ \mid\ \mathcal{N}(\hat{\mu}_{\text{out}}, \hat{\sigma}^2_{\text{out}}))}$
 \end{algorithmic}
 \caption{\textbf{VAR-LRT: LiRA on explanation variances.}
 The adversary trains shadow models on datasets with and without the target example, estimates parameters of the in- and out- distributions of sample variances of explanations (assuming Normal distributions of explanation variances), and runs a likelihood ratio test.
 }
 \label{alg:variancesattack}
\end{algorithm}

\textbf{Attacks on Explanation L1 and L2 Norms (L1-LRT/L2-LRT)} \quad \citet{shokri_white_box_attacks} previously highlighted disparities between gradient norm distributions of members and non-members, implying the efficacy of the \textit{gradient norm} as an attack statistic. Recently, \citet{wang2024pandoras} studied a white-box attack based on \textit{gradient norms} on open-source large language models. \textit{Explanation norms} are closely related to gradient norms, and we draw this connection---as well as explain our intuition behind constructing LiRAs based on explanation norms---in Appendix \ref{appendix:l1_l2_lrt}. \textbf{As far as we know, there is no prior work leveraging norms of model explanations in membership inference attack.} Algorithm \ref{alg:norm_lrt} in Appendix \ref{appendix:l1_l2_lrt} shows our explanation L1 norm-based LiRA algorithm (L1-LRT), which does so. The L2-LRT attack is almost identical but is based instead on L2 norms.

\textbf{Baselines} \quad Shokri et al.'s explanation-based attack, which we call the ``thresholding attack," is the main \textit{baseline attack} on which we improve. Moreover, in Section \ref{section:comparison_with_vanilla_lira}, we highlight that our attack methods perform competitively relative to the \textit{loss-based LiRA} baseline. This implies that our black-box methods are comparable to strong attacks that allow adversarial access to true labels.

\section{Experimental Results} \label{section:results}

\subsection{Setup} \label{subsection:setup}

We give full details on experimental setups and implementation in Appendix \ref{appendix:experimental_setup}, but in short, we fine-tune and report attack results on the following five datasets: CIFAR-10, CIFAR-100, Street View House Numbers (SVHN), Food 101, and German Traffic Sign Recognition Benchmark (GTSRB). For each dataset, we ``choose" a vision transformer model (out of 2-3 ImageNet pre-trained models analyzed per dataset) and hyperparameter setting to report in the main body, with additional and ablation experiment results in the appendices. Missing data in a few experimental setups is attributed to limitations in our compute resources. We release our code at \url{https://github.com/catherinehuang82/explaining-model-protecting-data}.

\subsection{Evaluation of the VAR-LRT Attack}
\begin{figure}[htb!]
    \centering

    \begin{subfigure}{\linewidth}
    \centering
\includegraphics[width=0.325\linewidth]{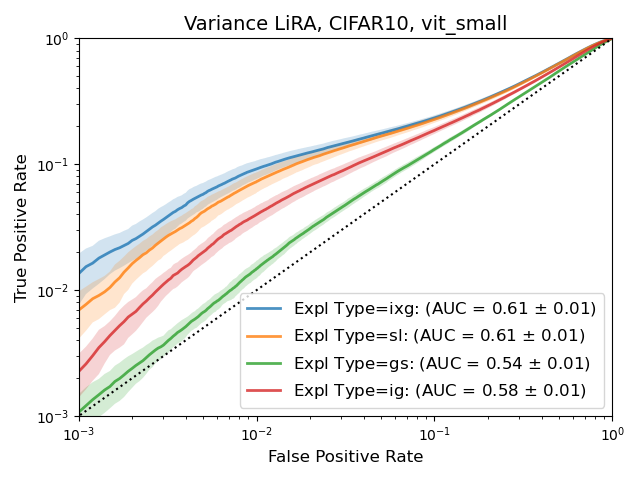}\includegraphics[width=0.325\linewidth]{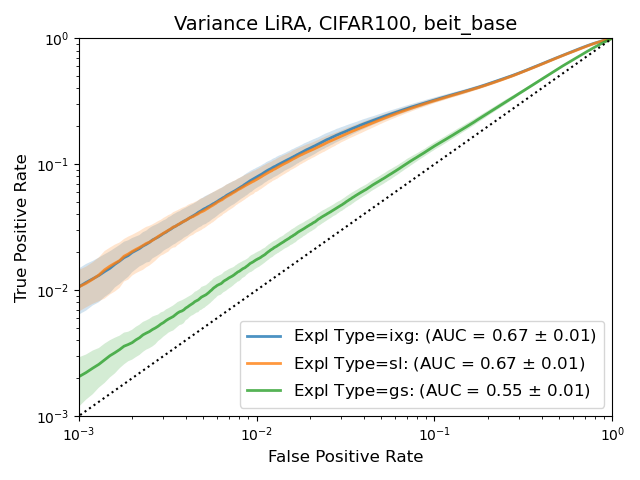} \includegraphics[width=0.325\linewidth]{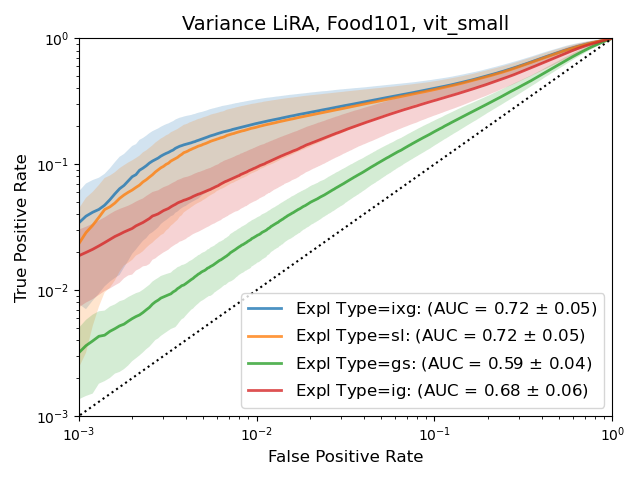}
\caption{VAR-LRT log-scaled ROC curves.}
\end{subfigure}

    \begin{subfigure}{\linewidth}
    \centering
\includegraphics[width=0.325\linewidth]{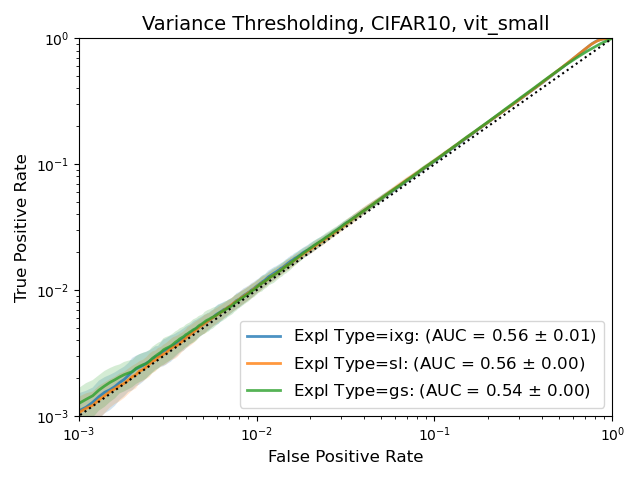}\includegraphics[width=0.325\linewidth]{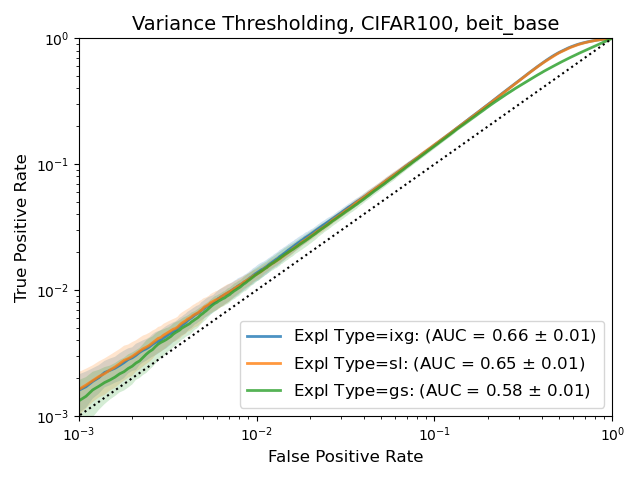}
\includegraphics[width=0.325\linewidth]{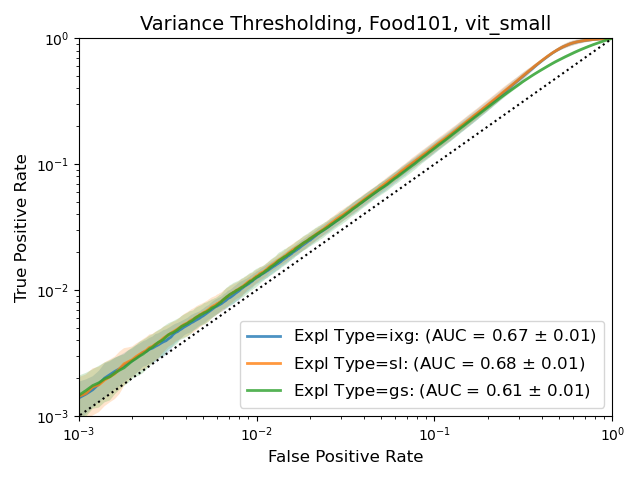}
\caption{Baseline thresholding attack log-scaled ROC curves.}
\end{subfigure}
    \caption{VAR-LRT vs. baseline thresholding attack ROCs for the CIFAR-10 (left), CIFAR-100 (middle), and Food 101 (right) datasets. We present results for all explanation methods under each dataset's chosen model and hyperparameter setting.}
    \label{fig:variance_nondp_all}
\end{figure}

We first present results on VAR-LRT and do an apples-to-apples comparison of this attack with Shokri et al.'s thresholding attack. Figure \ref{fig:variance_nondp_all} displays log-scaled ROC curves of the VAR-LRT versus baseline thresholding attacks for the CIFAR-$10$, CIFAR-$100$, and Food 101 datasets.

We observe from the log-scaled ROC curves in Figure \ref{fig:variance_nondp_all} that across datasets and explanation methods, VAR-LRT performs significantly better than random guessing at low FPRs. This means it confidently captures a small, known subset of training data members. Across datasets and explanation methods, VAR-LRT is more successful than the baseline attack at this task. More thoroughly, we present numerical results comparing VAR-LRT with the thresholding attack for four datasets and all explanation methods in Table \ref{tab:lira_vs_simple_variance}. We present each attack's performance on each metric and $\Delta$, the change between the two attacks' average performance. We encourage the reader to focus primarily on viewing the $\Delta$ columns but nonetheless provide complete metric values for reference. Almost all $\Delta$ values are green except for a few values for attacks on the GS explanation type, most of which correspond to the AUC metric---we explain in Appendix \ref{appendix:green_red_table_explanation} why these values do not undermine our conclusion throughout this section that VAR-LRT is a stronger attack than the thresholding attack.

\begin{table*}[h!]
\centering
\caption[Comparing explanation variance attack success.]{Comparing VAR-LRT vs. thresholding attack success. $\text{TPR}_{x}$ denotes the \text{TPR} at $\text{FPR} = x \ $ (i.e. $\text{FPR} = 100 \cdot x\%$). Green $\Delta$ values indicate metrics in which VAR-LRT has higher average value, and red $\Delta$ values indicate metrics in which VAR-LRT has lower average value. }

% \resizebox{0.72\textwidth}{!}{
% \begin{subtable}{\linewidth}
% \centering

\resizebox{0.72\linewidth}{!}{ %
\begin{tabular}{clcccccc}
% clcggcgg
\toprule
\multirow{2}{*}{Exp Type} & \multirow{2}{*}{Metric} & \multicolumn{3}{c}{CIFAR-$10$} &  \multicolumn{3}{c}{SVHN} \\
\cmidrule(l){3-5} \cmidrule(l){6-8}
& & \texttt{Thres.} & \texttt{VAR-LRT} & $\Delta$ \texttt{(Avg.)} & \texttt{Thres.} & \texttt{VAR-LRT} & $\Delta$ \texttt{(Avg.)}  \\
\cmidrule(lr){1-1} \cmidrule(lr){2-2} \cmidrule(l){3-5} \cmidrule(l){6-8}
\multirow{3}{*}{IXG} & $\text{TPR}_{.001}$ & $0.0012 \pm 0.0006$ & $0.0252 \pm 0.0084$ & \g{$0.0240$} & $0.0012 \pm 0.0001$ & $0.0121 \pm 0.0103$ & \g{$0.0471$} \\
& $\text{TPR}_{.01}$ & $0.0112 \pm 0.0021$ & $0.1195 \pm 0.0152$ & \g{$0.1083$} & $0.0099 \pm 0.0190$ & $0.0570 \pm 0.0241$ & \g{$0.0109$} \\
& AUC & $0.5588 \pm 0.0055$ & $0.6133 \pm 0.0093$ & \g{$0.0545$} & $0.5448 \pm 0.0080$ & $0.5863 \pm 0.0182$ & \g{$0.0415$}  \\
\cmidrule(lr){1-8}
\multirow{3}{*}{SL} & $\text{TPR}_{.001}$ & $0.0012 \pm 0.0006$ & $0.0154 \pm 0.0062$ & \g{$0.0142$} & $0.0012 \pm 0.0008$ & $0.0139 \pm 0.0114$ & \g{$0.0566$} \\
& $\text{TPR}_{.01}$ & $0.0111 \pm 0.0023$ & $0.1032 \pm 0.0171$ & \g{$0.0921$} & $0.0103 \pm 0.0029$ & $0.0668 \pm 0.0258$ & \g{$0.0126$} \\
& AUC & $0.5593 \pm 0.0050$ & $0.6082 \pm 0.0106$ & \g{$0.0489$} & $0.5456 \pm 0.0079$ & $0.5889 \pm 0.0181$ & \g{$0.0432$}  \\
\cmidrule(lr){1-8}
\multirow{3}{*}{IG} 
& $\text{TPR}_{.001}$ & $0.0012 \pm 0.0007$ & $0.0038 \pm 0.0020$ & \g{$0.0026$} & $0.0013 \pm 0.0007$ & $0.0046 \pm 0.0027$ & \g{$0.0152$} \\
& $\text{TPR}_{.01}$ & $0.0110 \pm 0.0022$ & $0.0573 \pm 0.0103$ & \g{$0.0464$} & $0.0107 \pm 0.0023$ & $0.0260 \pm 0.0071$ & \g{$0.0034$} \\
& AUC & $0.5539 \pm 0.0068$ & $0.5872 \pm 0.0193$ & \g{$0.0333$} & $0.5233 \pm 0.0051$ & $0.5412 \pm 0.0091$ & \g{$0.0180$}  \\
\cmidrule(lr){1-8}
\multirow{3}{*}{GS} 
 & $\text{TPR}_{.001}$ & $0.0016 \pm 0.0007$ & $0.0012 \pm 0.0007$ & \red{$0.0003$} & $0.0013 \pm 0.0008$ & $0.0024 \pm 0.0014$ & \g{$0.0047$} \\
& $\text{TPR}_{.01}$ & $0.0111 \pm 0.0019$ & $0.0181 \pm 0.0035$ & \g{$0.0069$} & $0.0103 \pm 0.0026$ & $0.0150 \pm 0.0036$ & \g{$0.0011$} \\
& AUC & $0.5404 \pm 0.0046$ & $0.5371 \pm 0.0056$ & \reddark{$-0.0033$} & $0.5229 \pm 0.0054$ & $0.5206 \pm 0.0064$ & \reddark{$-0.0023$}  \\
\bottomrule
% \end{tabular} 
% }
% \caption{CIFAR-10 and SVHN.}
    % \label{tab:subtable1}
  % \end{subtable}%
  % }

  % \vfill
  % \resizebox{0.72\textwidth}{!}{
% \begin{subtable}{\linewidth}
    % \centering
% \resizebox{0.72\columnwidth}{!}{ %
% \begin{tabular}{clcccccc}
% \toprule
\multirow{2}{*}{Exp Type} & \multirow{2}{*}{Metric} & \multicolumn{3}{c}{CIFAR-$100$} &  \multicolumn{3}{c}{Food 101} \\
\cmidrule(l){3-5} \cmidrule(l){6-8}
& & \texttt{Thres.} & \texttt{VAR-LRT} & $\Delta$ \texttt{(Avg.)} & \texttt{Thres.} & \texttt{VAR-LRT} & $\Delta$ \texttt{(Avg.)}  \\
\cmidrule(lr){1-1} \cmidrule(lr){2-2} \cmidrule(l){3-5} \cmidrule(l){6-8}
\multirow{3}{*}{IXG} 
& $\text{TPR}_{.001}$ & 
$0.0021 \pm 0.0010$ & $0.0200 \pm 0.0112$ & \g{$0.0179$} & $0.0012 \pm 0.0006$ & $0.0070 \pm 0.0018$ & \g{$0.0057$} \\
& $\text{TPR}_{.01}$ & 
$0.0158 \pm 0.0027$ & $0.1208 \pm 0.0271$ & \g{$0.1050$} & $0.0107 \pm 0.0021$ & $0.0225 \pm 0.0040$ & \g{$0.0118$} \\
& AUC & 
$0.6549 \pm 0.0100$ & $0.6708 \pm 0.0116$ & \g{$0.0157$} & $0.5106 \pm 0.0048$ & $0.5173 \pm 0.0050$ & \g{$0.0067$}  \\
\cmidrule(lr){1-8}
\multirow{3}{*}{SL} 
 & $\text{TPR}_{.001}$ & 
 $0.0018 \pm 0.0010$ & $0.0209 \pm 0.0109$ & \g{$0.0191$} & $0.0014 \pm 0.0007$ & $0.0021 \pm 0.0014$ & \g{$0.0062$} \\
& $\text{TPR}_{.01}$ & 
$0.0156 \pm 0.0029$ & $0.1176 \pm 0.0257$ & \g{$0.1020$} & $0.0106 \pm 0.0021$ & $0.0258 \pm 0.0041$ & \g{$0.0152$} \\
& AUC & 
$0.6522 \pm 0.0098$ & $0.6678 \pm 0.0112$ & \g{$0.0156$} & $0.5105 \pm 0.0043$ & $0.5170 \pm 0.0051$ & \g{$0.0066$}  \\
\cmidrule(lr){1-8}
\multirow{3}{*}{IG} 
& $\text{TPR}_{.001}$ & - & - & - & $0.0013 \pm 0.0009$ & $0.0028 \pm 0.0009$ & \g{$0.0015$} \\
& $\text{TPR}_{.01}$ & - & - & - & $0.0109 \pm 0.0026$ & $0.0147 \pm 0.0025$ & \g{$0.0039$} \\
& AUC & - & - & - & $0.5065 \pm 0.0048$ & $0.5074 \pm 0.0047$ & \g{$0.0009$}  \\
\cmidrule(lr){1-8}
\multirow{3}{*}{GS}
 & $\text{TPR}_{.001}$ & $0.0019 \pm 0.0010$ & $0.0027 \pm 0.0011$ & \g{$0.0008$} & $0.0012 \pm 0.0007$ & $0.0013 \pm 0.0007$ & \g{$0.0002$} \\
& $\text{TPR}_{.01}$ & $0.0152 \pm 0.0019$ & $0.0200 \pm 0.0036$ & \g{$0.0049$} & $0.0110 \pm 0.0022$ & $0.0116 \pm 0.0023$ & \g{$0.0006$} \\
 & AUC & $0.5847 \pm 0.0065$ & $0.5572 \pm 0.0080$ & \reddark{$-0.0275$} & $0.5057 \pm 0.0052$ & $0.5021 \pm 0.0033$ & \reddark{$-0.0036$}  \\
\bottomrule
\end{tabular}
}
% \caption{CIFAR-100 and Food 101.}
    \label{tab:subtable1}
  % \end{subtable}%

  % }
\label{tab:lira_vs_simple_variance}
\end{table*}

\subsection{Evaluation of the L1-LRT and L2-LRT Attacks}

\begin{figure}[htb!]
    \centering

    \begin{subfigure}{\linewidth}
    \centering
\includegraphics[width=0.325\linewidth]{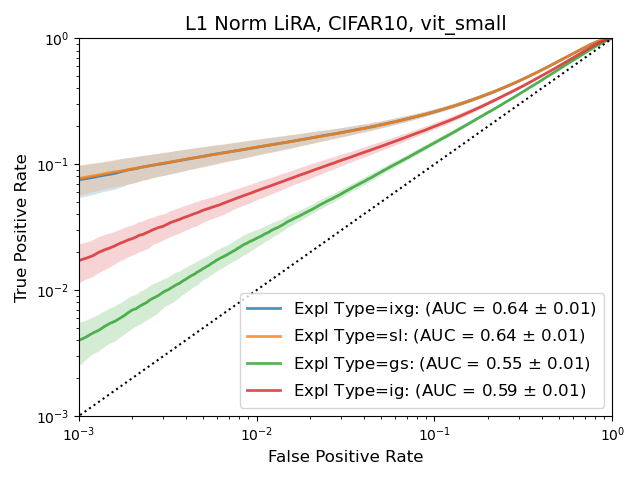}\includegraphics[width=0.325\linewidth]{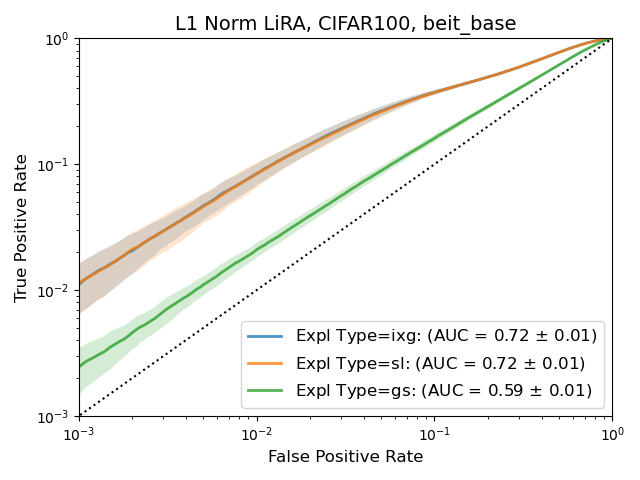}
\includegraphics[width=0.325\linewidth]{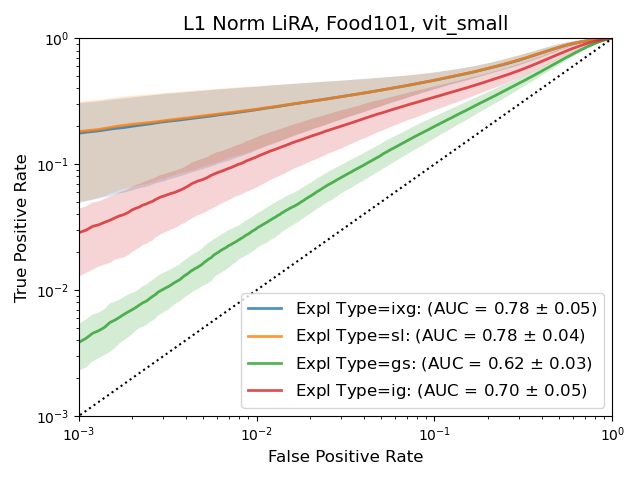}
\caption{L1-LRT log-scaled ROC curves.}
\end{subfigure}

    \begin{subfigure}{\linewidth}
    \centering
\includegraphics[width=0.325\linewidth]{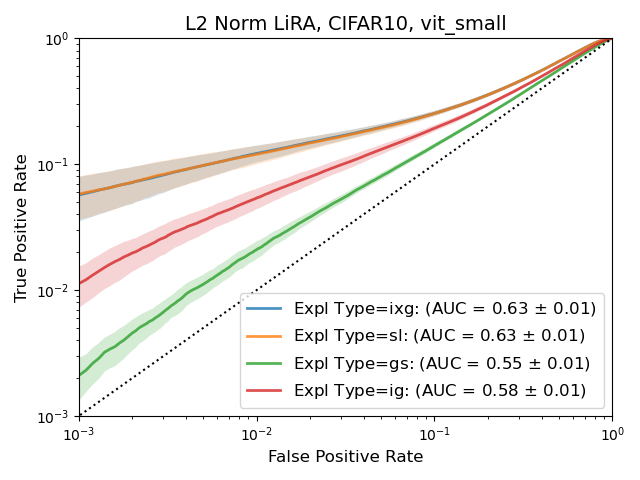}\includegraphics[width=0.32\linewidth]{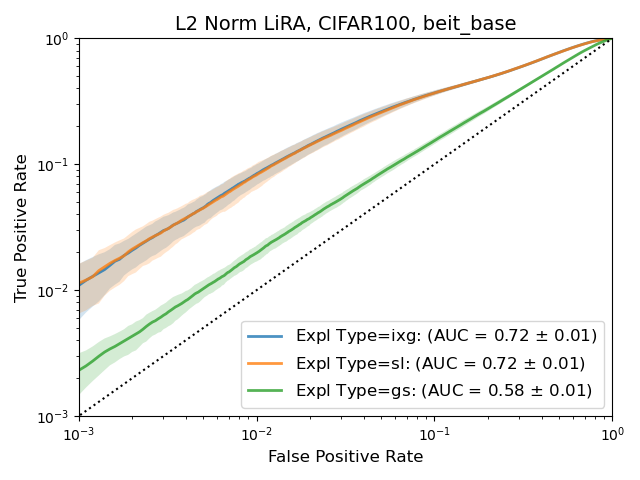}
\includegraphics[width=0.325\linewidth]{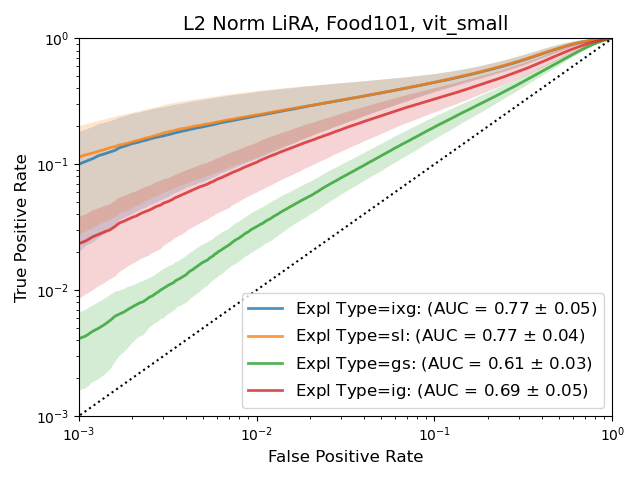}
\caption{L2-LRT log-scaled ROC curves.}
\end{subfigure}
    
    \caption{L1-LRT and L2-LRT attack results for the CIFAR-10 (left), CIFAR-100 (middle), and Food 101 (right) datasets.}
    \label{fig:l1_l2_norm_nondp_all}
\end{figure}

Figure \ref{fig:l1_l2_norm_nondp_all} displays L1-LRT and L2-LRT attack ROCs for the CIFAR-$10$, CIFAR-$100$, and Food 101 datasets. L1-LRT and L2-LRT behave similarly to one another across the ROC curve, and both are highly successful, objectively and even relative to VAR-LRT. (We see that L1-LRT generally performs better than L2-LRT and hypothesize why in Appendix \ref{appendix:l1_lrt_vs_l2_lrt}.) To further highlight our most successful attack, L1-LRT, Table \ref{tab:l1_lrt_nondp_table} shows numerical L1-LRT results for all five datasets; we see many bolded quantities highlighting where TPR at FPR $= x$ is at least $10 \cdot x$. We also observe that \textbf{across the whole table (every setting), the mean value of TPR at FPR $ = x$ is higher than $x$}. This means that attacks perform reliably across the board, and a substantial number of attacks---especially on IXG and SL explanation types---perform exceedingly well at small FPR values.

\begin{table*}[htb!]
\caption[Numerical attack results for L1-LRT, all datasets.]{Numerical attack results for L1-LRT. The bolded quantities show where TPR at FPR = $x$ is at least $10 \cdot x$.}
\centering
\resizebox{0.72\textwidth}{!}{ %
\begin{tabular}{clccccc}
\toprule
\multirow{2}{*}{Exp Type} & \multirow{2}{*}{Metric} & \multicolumn{5}{c}{Dataset}\\
\cmidrule(l){3-7}
& & \texttt{CIFAR-$10$} & \texttt{CIFAR-$100$} & \texttt{Food 101} & \texttt{SVHN} & \texttt{GTSRB} \\
\cmidrule(lr){1-1} \cmidrule(lr){2-2} \cmidrule(l){3-7}
\multirow{3}{*}{IXG} 
& $\text{TPR}_{.001}$ &  $\boldsymbol{0.093} \pm 0.022$ &
$\boldsymbol{0.022} \pm 0.013$ &
$\boldsymbol{0.203} \pm 0.140$ & 
$\boldsymbol{0.015} \pm 0.014$ &
$0.009 \pm 0.002$ \\
& $\text{TPR}_{.01}$ & 
$\boldsymbol{0.156} \pm 0.018$ &
$\boldsymbol{0.130} \pm 0.037$ &
$\boldsymbol{0.310} \pm 0.131$ & 
$0.065 \pm 0.029$ &
$0.027 \pm 0.004$ \\
& AUC &  
$0.639 \pm 0.008$ &  
$0.716 \pm 0.012$ & 
$0.780 \pm 0.046$ & 
$0.603 \pm 0.018$ &
$0.518 \pm 0.005$ \\
\cmidrule(lr){1-7}
\multirow{3}{*}{SL}
 & $\text{TPR}_{.001}$ & 
$\boldsymbol{0.093} \pm 0.022$ &  
$\boldsymbol{0.021} \pm 0.011$ &
 $\boldsymbol{0.210} \pm 0.143$ & 
  $\boldsymbol{0.017} \pm 0.015$ &
  $\boldsymbol{0.012} \pm 0.002$ \\
& $\text{TPR}_{.01}$ & 
$\boldsymbol{0.155} \pm 0.019$ &  
$\boldsymbol{0.128} \pm 0.035$ & 
$\boldsymbol{0.309} \pm 0.132$ & 
$0.077 \pm 0.030$ & 
$0.030 \pm 0.005$ \\
 & AUC & 
 $0.639 \pm 0.009$ & 
 $0.716 \pm 0.011$ &
 $0.782 \pm 0.043$ &
 $0.605 \pm 0.018$ & 
 $0.518 \pm 0.005$ \\
\cmidrule(lr){1-7}
\multirow{3}{*}{IG} 
& $\text{TPR}_{.001}$ &  $\boldsymbol{0.026} \pm 0.008$ &
$-$ &
$\boldsymbol{0.044} \pm 0.024$ &  
$0.006 \pm 0.003$ & 
$0.004 \pm 0.001$\\
& $\text{TPR}_{.01}$ &
$0.080 \pm 0.012$ & 
$-$ &
$\boldsymbol{0.159} \pm 0.032$ & 
$0.008 \pm 0.007$ & $0.017 \pm 0.003$  \\
& AUC & 
$0.590 \pm 0.009$ &
$-$ & 
$0.700 \pm 0.051$  & 
$0.554 \pm 0.009$ & 
$0.508 \pm 0.004$ \\
\cmidrule(lr){1-7}
\multirow{3}{*}{GS} 
 & $\text{TPR}_{.001}$ &  
 $0.006 \pm 0.002 $ & 
 $0.003 \pm 0.001$ &
 $0.006 \pm 0.003$ &  
 $0.003 \pm 0.001$ & 
 $0.002 \pm 0.001$ \\
& $\text{TPR}_{.01}$ &  
$0.033 \pm 0.005$ & 
$0.027 \pm 0.005$ &
$0.044 \pm 0.016$ & 
$0.017 \pm 0.004$ & 
$0.012 \pm 0.003$  \\
& AUC &  
$0.554 \pm 0.006$ & 
$0.586 \pm 0.009$ &
$0.616 \pm 0.032$ & 
$0.532 \pm 0.007$ & 
$0.502 \pm 0.004$ \\
\bottomrule
\end{tabular} 
}
\label{tab:l1_lrt_nondp_table}
\end{table*}

\textbf{More Non-Private Attack Results and Ablation Experiments} \quad More results on model and attack performance and ablation experiments are in Appendices \ref{appendix:model_performance}, \ref{appendix:additional_var_lrt}, \ref{appendix:additional_l1_l2_lrt}, 
and \ref{appendix:ablation_experiments}.

\subsection{Comparison with Loss-Based LiRA}\label{section:comparison_with_vanilla_lira}

Table \ref{tab:comparison_with_loss_lira} compares our attacks to the LiRA baseline based on per-example cross-entropy loss (``Loss LiRA"). L1-LRT and L2-LRT dominate Loss LiRA across all reported metrics, and VAR-LRT is also competitive. Given that Loss LiRA is a strong attack to begin with among attacks that allow adversarial access to true labels (as explained in Section \ref{section:preliminaries} and shown in \citet{carlini}), this result implies further privacy risk: our methods perform competitively even with black-box limitations on what the adversary has access to. This is a brief proof of concept---further work is helpful for benchmarking our attacks against attacks based on other, possibly more ``traditional" signals, such as loss.

\def\thickhline{\noalign{\hrule height.8pt}}

\begin{table}[htb!]
\centering
\caption{Attack metrics compared with the loss LiRA attack on CIFAR-10, IXG explanations, \texttt{vit\_small} model. Bold quantities indicate TPR @ low FPR metrics where our methods outperform loss LiRA.}
\vspace{0.1in}
\resizebox{0.9\columnwidth}{!}{
 \begin{tabular}{r|ccc}
  \toprule
    Attack  & $\text{TPR}_{.001}$ & 
     $\text{TPR}_{.001}$ & AUC  \\\hline
  Loss LiRA & $0.054 \pm 0.007$ & $0.095 \pm 0.012$ & $0.570 \pm 0.009$ \\ \hdashline
  L1-LRT & $\boldsymbol{0.093} \pm 0.022$ & $\boldsymbol{0.156} \pm 0.018$ & $0.639 \pm 0.008$ \\
  L2-LRT & $\boldsymbol{0.078} \pm 0.021$ & $\boldsymbol{0.146} \pm 0.016$ & $0.633 \pm 0.007$ \\
  
  VAR-LRT & $0.025 \pm 0.008$ & $\boldsymbol{0.120} \pm 0.015$ & $0.613 \pm 0.009$ \\
  \bottomrule
 \end{tabular}
 }
 \label{tab:comparison_with_loss_lira}
\end{table}

\subsection{Mitigating Attack Success with Differential Privacy}\label{section:dp_mitigation}

We investigate whether \citet{bu2023automatic}'s accuracy-optimizing differentially private fine-tuning can successfully mitigate the success of these explanation-based attacks while preserving model accuracy. We approach this part by treating attack results on non-privately fine-tuned models as \textit{baseline results} upon which we wish to improve, now aiming for \textit{lower} attack success metrics. 

\begin{figure}[htb!]
    \centering
    \begin{subfigure}{\linewidth}
    \centering
    \includegraphics[width=0.325\linewidth]{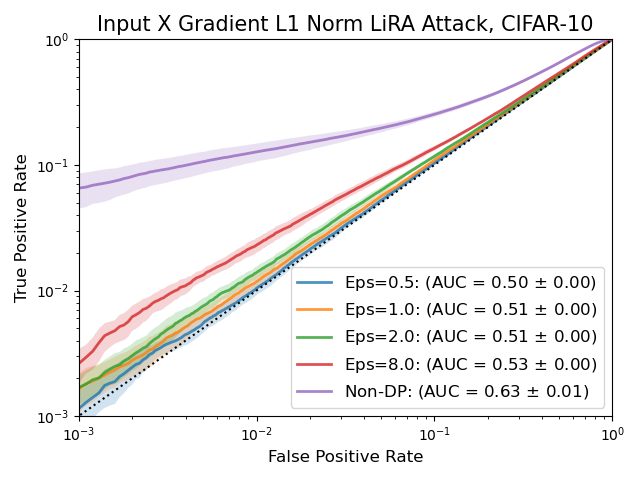}\includegraphics[width=0.325\linewidth]{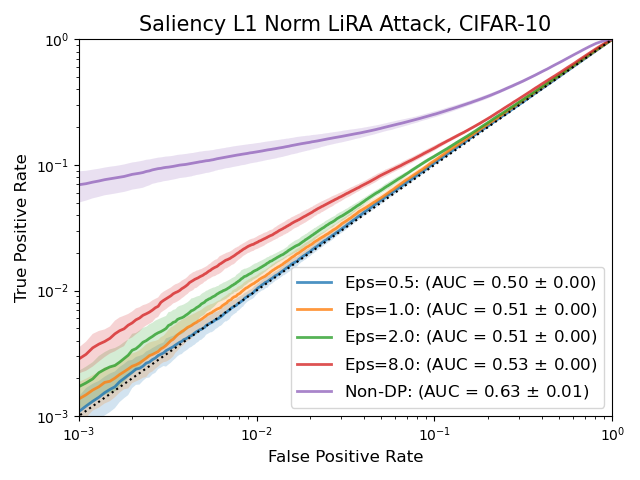}
\includegraphics[width=0.325\linewidth]{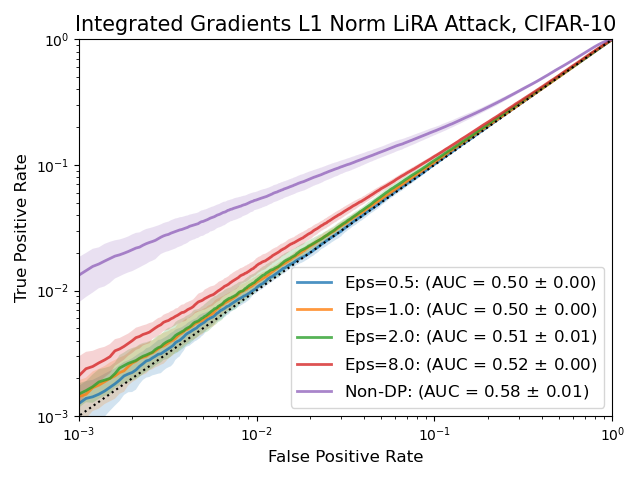}
    \caption{CIFAR-10 (IXG, SL, IG).}
    \end{subfigure}

    \vfill
    \begin{subfigure}{\linewidth}
    \centering
    \includegraphics[width=0.325\linewidth]{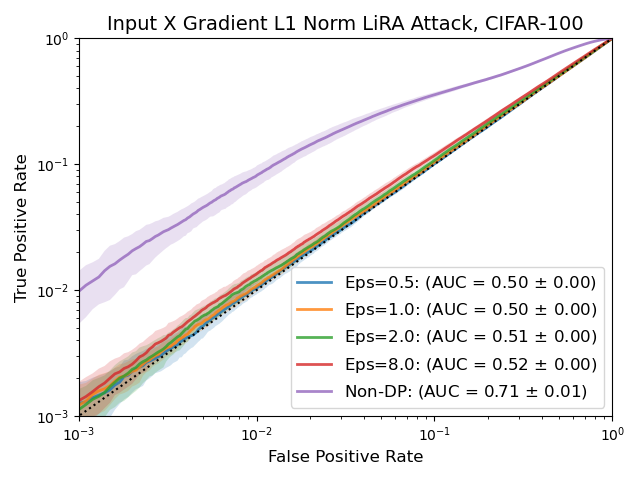}\includegraphics[width=0.325\linewidth]{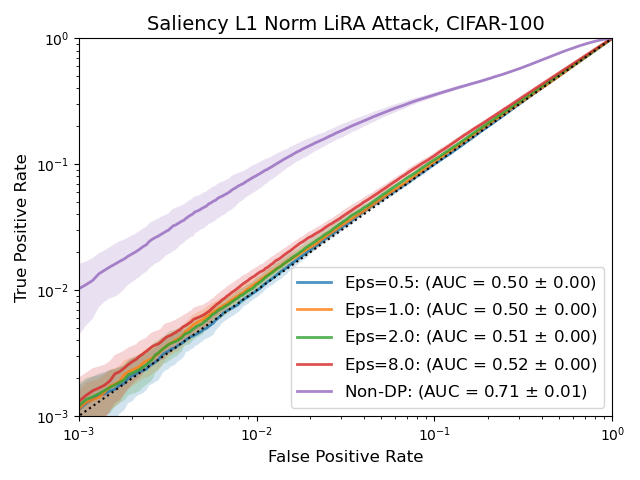}
    \caption{CIFAR-100 (IXG, SL).}
    \end{subfigure}

    % \vfill
    % \begin{subfigure}{\linewidth}
    % \centering
    % \includegraphics[width=0.32\linewidth]{figures/norms_dp_svhn/Input X Gradient L1 Norm LiRA Attack, 50 Epochs, SVHN.png} \includegraphics[width=0.32\linewidth]{figures/norms_dp_svhn/Saliency L1 Norm LiRA Attack, 50 Epochs, SVHN.png} 
    % \includegraphics[width=0.32\linewidth]{figures/norms_dp_svhn/Integrated Gradients L1 Norm LiRA Attack, 50 Epochs, SVHN.png} 
    % \caption{SVHN (IXG, SL, IG).}
    % \end{subfigure}
    
    \caption[L1-LRT attack success of non-private vs. DP fine-tuned models.]{L1-LRT attack success of non-private (purple curves) vs. DP fine-tuned models (other curves). We show one plot per explanation method: IXG (left), SL (middle), and (with the exception of CIFAR-100) IG (right). Each subplot shows curves for $\epsilon = 0.5, 1.0, 2.0, 8.0$.}
    \label{fig:l1_norm_dp_all}
\end{figure}

Figure \ref{fig:l1_norm_dp_all} shows ROC curves highlighting the impact of DP fine-tuning on our most powerful attack, L1-LRT. We observe that in each subplot---that is, across datasets and explanation types---the ROC curves corresponding to models fine-tuned with DP hug the ``random guessing" diagonal line much more closely than the baseline ROC curves do. With CIFAR-$10$, the $\varepsilon = 8.0$ setting (and, for IXG and SL explanations, the $\varepsilon=2.0$ setting) gives slight attack success. However, the other curves associated with DP fine-tuning in Figure \ref{fig:l1_norm_dp_all} show attacks that behave no better than random guessing, both on average and in low FPR.

DP fine-tuned models give rise to explanation-based attacks that certainly do not demonstrate privacy risk at the level that the non-private attacks do. Thus, while we cannot guarantee that explanations that come from DP models will always deliver completely unsuccessful membership inference attacks, we still strongly conclude that DP fine-tuning substantially diminishes explanation-based MIA success to the point of doing only minimally better than ``random guessing." Appendix \ref{appendix:more_dp_results} shows that this conclusion holds across the other attack types, and Appendix \ref{appendix:dp_model_performance} shows, importantly, that DP mitigates attack success \textit{while maintaining model accuracy}.

\section{Discussion} \label{section:future}
The lack of trust in post-hoc explanations contributes to the slow adoption of machine learning in high-stakes domains. This paper reveals unforeseen vulnerabilities of feature attribution explanations to membership inference by introducing two new attacks that respectively leverage sample variances and norms of attribution vectors. These attacks are significantly more successful than existing attacks that leverage explanations, particularly at confidently identifying specific training set members.

We also show that when foundation models are fine-tuned with differential privacy, post-hoc explanations yield significantly lower explanation-based attack success compared to when models are fine-tuned without DP---all while preserving model accuracy. Our work patches a gap in literature---there is no prior empirical work in deep learning that thoroughly quantifies the relationship between DP training and the subsequent privacy risks of a broad class of post-hoc model explanations.

Within adversarial ML, there remain open research directions concerning how post-hoc model explanations may be leveraged to compromise data privacy. For example, can explanations be leveraged \textit{alongside} traditional signals (e.g. loss, confidence) to compromise privacy to a greater extent? Can we get similar attack success without requiring adversarial training of shadow models? Can we formally quantify attack success (i.e. TPR at certain FPR)? How successful might other non-MIA attack types that leverage explanations be?

Keeping trade-offs in mind, we emphasize that there remains a barrier between the idea of differential privacy as a promising privacy risk mitigation approach, and its tangible deployment in contexts that necessitate model explanations. The latter entails adequate analysis on if privacy sacrifices on the \textit{quality} or \textit{usefulness} of explanations---this is out of the scope of this work but remains necessary to be studied.

\newpage

\bibliography{example_paper}
\bibliographystyle{icml2024}

%%%%%%%%%%%%%%%%%%%%%%%%%%%%%%%%%%%%%%%%%%%%%%%%%%%%%%%%%%%%%%%%%%%%%%%%%%%%%%%
%%%%%%%%%%%%%%%%%%%%%%%%%%%%%%%%%%%%%%%%%%%%%%%%%%%%%%%%%%%%%%%%%%%%%%%%%%%%%%%
% APPENDIX
%%%%%%%%%%%%%%%%%%%%%%%%%%%%%%%%%%%%%%%%%%%%%%%%%%%%%%%%%%%%%%%%%%%%%%%%%%%%%%%
%%%%%%%%%%%%%%%%%%%%%%%%%%%%%%%%%%%%%%%%%%%%%%%%%%%%%%%%%%%%%%%%%%%%%%%%%%%%%%%
\newpage
\appendix
\onecolumn

\section{Appendix}

Our appendices are organized into the following parts:
\begin{itemize}
    \item \ref{appendix:vision_transformers}: The Case for Foundation Models and Vision Transformers
    \item \ref{appendix:post_hoc_explanations}: Post-Hoc Feature Attribution Explanations
    \item \ref{appendix:more_on_dp}: More on Differential Privacy, DP-SGD, and DP with Automatic Gradient Norm Clipping
    \item \ref{appendix:l1_l2_lrt}: L1-LRT/L2-LRT Intuition and Algorithm
    \item \ref{appendix:experimental_setup}: Experimental Setups and Implementation Details
    \item \ref{appendix:model_performance}: Performance of Non-Private Models
    \item \ref{appendix:additional_var_lrt}: More Non-Private VAR-LRT Results
    \item \ref{appendix:additional_l1_l2_lrt}: More Non-Private L1-LRT/L2-LRT Results
    \item \ref{appendix:ablation_experiments}: Non-Private Ablation Experiments
    \item \ref{appendix:dp_model_performance}: Performance of Models Fine-Tuned with Differential Privacy
    \item \ref{appendix:more_dp_results}: More Results on the Impact of Differentially Private Fine-Tuning on Attack Success
\end{itemize}

\newpage

\section{The Case for Foundation Models and Vision Transformers}\label{appendix:vision_transformers}
Fine-tuning happens when a pre-trained foundation model is then trained on a smaller, more specific new task \cite{vit}. Foundation models in general are desirable for a variety of reasons with respect to our research questions:
\begin{itemize}
    \item Foundation models are generally trained on data with public access, which means the models do not touch sensitive data (until possibly during downstream tasks). Hence, pre-trained foundation models adhere to the right to privacy.
    \item Foundation models perform remarkably well with complex problems on complex datasets. They are less prone to overfitting, having been pre-trained on broad data and able to effectively handle a wide range of inputs. Foundation models' architectures have large size, depth, and scale that allow for their state-of-the-art quality.
    \item Foundation models such as large language models (e.g. GPT \cite{openai2024gpt4}) and vision-language models (e.g. CLIP \cite{radford2021learning}) are widely applied to complex real-world settings. In terms of vision tasks, for example, foundation models are used for applications ranging from medical imaging \cite{azad2023foundational, PMID:37045921, sowrirajan2021mococxr, Ke_2021, tian2024unigradicon, wu2023generalist} to astronomy \cite{grezes2021building, nguyen2023astrollama} to robotics \cite{kawaharazuka2024realworld, brohan2023rt2}. By virtue of their versatility and widespread use, foundation models are thus a viable choice of model type in settings involving sensitive personal data.
    \item Fine-tuning foundation models on downstream tasks requires substantially fewer computational resources than training the model from scratch. Fine-tuning generally requires fewer epochs than standard training. To conduct membership inference in this work, we fine-tune tens of shadow models per attack experiment, and doing so is computationally more feasible than training these numerous large models from scratch.
\end{itemize}

For these reasons, in particular for the first reason on privacy defense, we choose to evaluate our methods using large pre-trained foundation models on fine-tuned tasks.

\subsection{The Vision Transformer Architecture}
In this work, we evaluate our methods on image classification tasks. We choose to focus on image classification tasks rather than text classification, since post-hoc explanations are conceptually better defined for images: each pixel is a feature, and post-hoc explanations reveal which pixels in an image are most influential to a model's prediction of that image's class. Text corpora are typically higher-dimensional and less standardized than images (which can readily be scaled to a fixed, standardized dimension), meaning that explanations on text-datasets are especially sparse. Furthermore, flagship papers on post-hoc explainability methods \cite{shap, sundararajan17a, shrikumar} typically evaluate their explanation methods on image data. Such image classification tasks are commonly trained with convolutional neutral network (CNN) foundation models; for example, \citet{he2015deep} famously introduced the state-of-the-art residual network (``ResNet") CNN architecture.

Separately, in the natural language processing domain, the Transformer architecture was proposed by Vaswani et al. \citet{attention_is_all_you_need} for machine translation. Since then, Transformer-based architectures have become state-of-the-art in many NLP tasks. Transformers rely on a \textit{self-attention} mechanism that is scalable, efficient, and captures both short-term and long-term dependencies among text sequences. Compared with previously prevalent CNN and recurrent neural network approaches to NLP tasks, Transformers have both higher performance and higher speed. Transformers are commonly pre-trained on large text corpora and then fine-tuned on smaller, more specific tasks, making them a suitable foundation model. 

Applying Transformers to image classification tasks naively would require that each pixel attend to each other pixel; this is intractable. \citet{vit} propose the state-of-the-art solution in the \textit{vision transformer (ViT)} architecture: the ViT reshapes each original input image $\mathbf{x} \in \mathbb{R}^{H \times W \times C}$ into a sequence of flattened $2$D patches $\mathbf{x}_p \in \mathbb{R}^{N \times (P^2 \cdot C)}$, where $(H \times W)$ are the dimensions of the original image, $C$ is the number of (color) channels, $(P \times P)$ are the dimensions of each image patch, and $N = \frac{HW}{P^2}$ is the number of patches. After each image is split into fixed-size patches, ViT linearly embeds each image and adds positional embeddings to incorporate positional information (of the patches within each image). The embedded vectors are then fed into ViT's Transformer encoder, which is built with alternating layers of multiheaded self-attention units and multilayer perception (MLP) units. Each MLP block contains two layers with a Gaussian Error Linear Unit (GELU) activation function \cite{vit}. The GELU function is a high-performing neural network activation function that often yields a performance improvement upon the more vanilla ReLU activation function \cite{gelu}. For $Z \sim \mathcal{N}(0,1)$ a Standard Normal random variable, GELU is defined as
\[\text{GELU}(x) = x P(Z \le x).\] 

\begin{figure}[h!]
\begin{center}
    \includegraphics[width=2.3in]{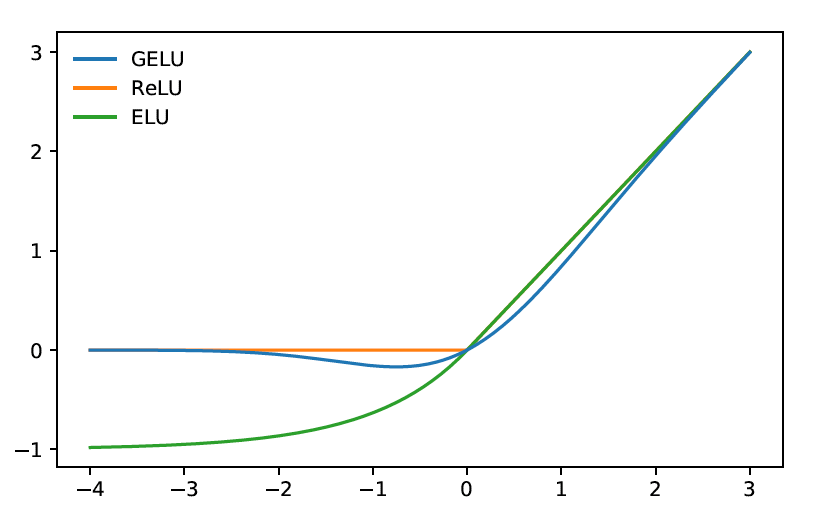}
    \caption[The GELU, ReLU, and ELU (Exponential Linear Unit) activation functions.]{The GELU, ReLU, and ELU (Exponential Linear Unit) \cite{elu} activation functions. The vision transformer architecture uses GELU activations.}
    \label{fig:gelu}
    \end{center}
\end{figure}

\begin{figure}[h!]
\begin{center}
    \includegraphics[width=4.1in]{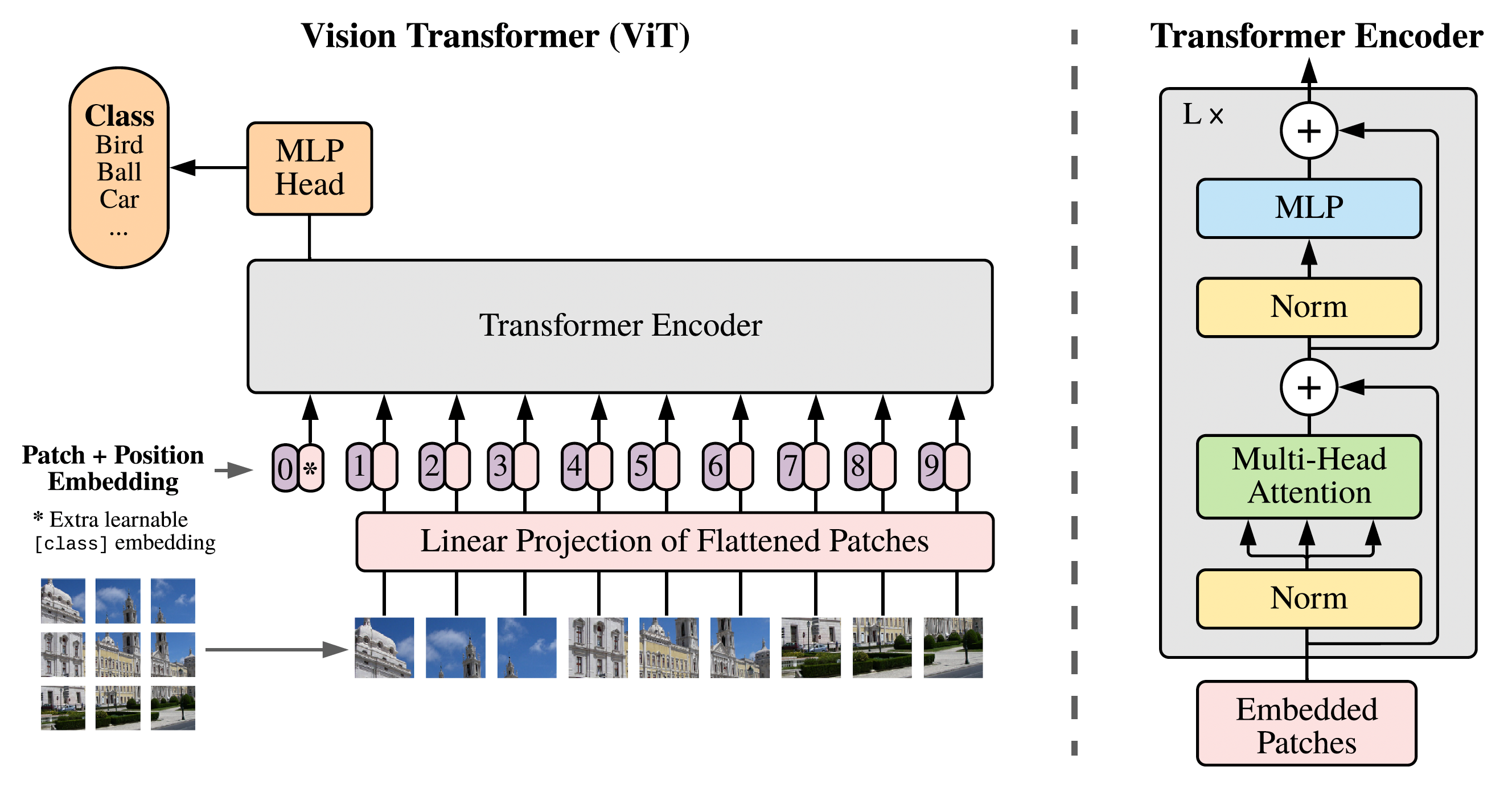}
    \caption[Overview of the vision transformer (ViT) model architecture.]{Overview of the vision transformer (ViT) model architecture. ViT splits an image into patches, embeds them (linearly and positionally), and feeds the embeddings into a Transformer encoder.}
    \label{fig:vitmodeloverview}
    \end{center}
\end{figure}

Figure \ref{fig:gelu} visualizes GELU compared with other common neural network activation functions. Figure \ref{fig:vitmodeloverview} provides an overview of the vision transformer architecture.

In this work, we use vision transformers in our experiments because not only are vision transformers state-of-the-art foundation models, but they perform better on fine-tuned downstream datasets than ResNet.

\section{Post-Hoc Feature Attribution Explanations}\label{appendix:post_hoc_explanations}

We study the following backpropagation-based feature attribution methods: Input $*$ Gradient (IXG), Saliency Maps (SL), Integrated Gradients (IG), and (a gradient-based approximation to) SHAP (GS). We describe each of these methods, as well as any desirable properties they exhibit (according to the pioneers of these methods).

\subsection{Input $*$ Gradient (\texttt{IXG})}
We first introduce the Input * Gradient technique \cite{shrikumar}. This attribution vector is relatively simple to generate, computed by taking the partial derivatives of the output with respect to each input feature and multiplying them with the input itself: \[\varphi_i(\mathbf{x}) = x_i \cdot \frac{\partial f_{\theta}(\mathbf{x})}{\partial x_i}.\]

\subsection{Saliency Maps (\texttt{SL})}
Saliency maps are almost equivalent to taking vanilla gradients. The only difference is that this method computes the \textit{absolute value} of the gradient with respect to each input feature.

The interpretation of absolute value is that features with the highest absolute gradient need to be perturbed the least in order for the model's predicted output to change the most. The ensuing limitation of saliency maps is that they do not differentiate between features that contribute positively to prediction and features that contribute negatively. However, since our explanation-based attack methods are based on scalar summaries of attribution vectors (e.g. variance and norms), this limitation is not a direct concern.

\subsection{Integrated Gradients (\texttt{IG})}\label{section:integrated_gradients}
Integrated gradients \cite{sundararajan17a} similarly computes the partial derivatives of the output with respect to each input feature. However, instead of only computing the gradient on the original input, IG computes the \textit{average} gradient as the input varies along a linear path from a baseline $\mathbf{x}_{BL}$ to $\mathbf{x}$ (usually, $\mathbf{x}_{BL} = \boldsymbol{0}$). The mathematical definition of IG is \[\varphi_{IG}(\mathbf{x})_i = (x_i - x_{BL,i}) \cdot \int_{\alpha = 0}^{1}\frac{\partial f_{\theta}(\mathbf{x}^{\alpha})}{\partial x_i^{\alpha}} d\alpha \bigg\rvert_{\mathbf{x}^{\alpha}= \mathbf{x} + \alpha(\mathbf{x} - \mathbf{x}_{BL})}.\]

Through an axiomatic approach, IG is designed to satisfy three desirable properties of attribution methods: sensitivity, implementation invariance, and completeness. Implementation-wise, we can only approximate the integral by taking a Riemann sum over a discrete number of gradients along the linear path from baseline to input.

\textbf{Sensitivity} \quad Sensitivity means that given a point $\mathbf{x} \in \mathcal{X}$ such that $x_i \ne x_{BL, i}$ and $f_{\theta}(\mathbf{x}) \ne f_{\theta}(\mathbf{x}_{BL})$, then $\varphi_i(\mathbf{x}) \ne 0$. In words, sensitivity asserts that for every input and baseline that differ in one feature but have different predictions, then the explanation method should give that feature a non-zero attribution.

\textbf{Completeness} \quad Completeness means that $\sum_{i=1}^{n} \varphi_i(\mathbf{x}) = f_{\theta}(\mathbf{x}) - f_{\theta}(\mathbf{x}_{BL})$: the attributions sum up to the difference between the output of $f_{\theta}$ at the input $\mathbf{x}$ and the baseline $\mathbf{x}_{BL}$.

\textbf{Implementation invariance} \quad Two models $f_1$ and $f_2$ are functionally equivalent if $f_1(\mathbf{x}) = f_2(\mathbf{x})$ for all inputs $\mathbf{x}$. The implementation invariance axiom asserts that explanations should be identical for functionally equivalent models. 

\subsection{SHapley Additive exPlanations (SHAP; Abbreviated as GS)}\label{section:gradient_shap}

In the original SHAP paper, \citet{shap} assume an additive \textit{explanation model} $g$: $g$ is an interpretable approximation of the original model $f_{\theta}$ that is a linear combination of binary variables. The authors show that only one possible additive explanation model $g$ satisfies the three axiomatic properties of local accuracy, missingness, and monotonicity (details of which are beyond the scope of this work). Further, the corresponding feature attribution values $\phi_i$ of model $g$ coincide with Shapley values \yrcite{shapley} in cooperative game theory. The SHAP explanation framework is based on the Shapley values of a conditional expectation function derived from $f_{\theta}$, the details of which are beyond the scope of this work. At a high level, SHAP values set $\varphi_i$ to the change in the expected model prediction when conditioning on feature $i$. 

SHAP values are difficult to compute exactly, and in this work, we use a \textit{gradient-based approximation} to SHAP values that approximates the expectation of gradients * (inputs - baselines) \cite{captum}. The approximation works as follows: we add Gaussian random noise to each input sample multiple times, select random points along the path between the input $\mathbf{x}$ and a baseline $\mathbf{x}_{BL},$ and compute the gradient of outputs with respect to these points on the path. We use this approximation and refer to it as ``Gradient SHAP" with abbreviation ``GS."

\section{More on Differential Privacy, DP-SGD, and DP with Automatic Gradient Norm Clipping} \label{appendix:more_on_dp}

Formally, a randomized mechanism $\mathcal{M}$ with domain $\mathcal{D}$ and range $\mathcal{R}$ satisfies \textbf{$(\varepsilon, \delta)$-Differential Privacy} ($(\varepsilon, \delta)$-DP) if for any two adjacent input datasets $D, D' \sim \mathbb{D}$ differing by one row, and any subset of outputs $S \subseteq \mathcal{R}$, $ Pr[\mathcal{M}(D) \in S] \le e^{\varepsilon} \cdot Pr[\mathcal{M}(D') \in S] + \delta$. The lower the $\varepsilon$ (the privacy parameter), the stronger the privacy protection. A value of $\varepsilon = \infty$ represents that a mechanism lacks privacy. 

\textbf{DP Stochastic Gradient Descent (DP-SGD)} \quad \citet{abadi} propose a differentially private stochastic gradient descent (DP-SGD) mechanism to train models with DP. It works by adding Gaussian distributed random noise (with standard deviation $\sigma$) to gradients. Higher $\sigma$ (i.e. adding random noise with higher variance) leads to lower $\varepsilon$. 

Algorithm \ref{alg:dpsgd} outlines Abadi et al.'s method for training a model with parameters $\theta$. Before adding noise to gradients, DP-SGD upper-bounds the norm of the gradient to be within $C$ via gradient norm clipping; this is to enforce that the sensitivity of the gradients be exactly $C$ (instead of being unbounded). DP-SGD then involves adding $\mathcal{N}(\mathbf{0}, \sigma^2 C^2 \mathbf{I})$ distributed random noise, where $\sigma$ is a pre-specified noise scale. Abadi et al. use the \textit{moments accountant} privacy loss budgeting technique and the Gaussian Mechanism to prove that for appropriately chosen $\sigma$, DP-SGD is $(O(q\varepsilon \sqrt{T}), \delta)$-DP, where $q$ is the sampling ratio per batch. 

\begin{algorithm}[t]
	\caption{Differentially Private SGD (Outline)}\label{alg:dpsgd}
	\begin{algorithmic}
	\REQUIRE Examples $\{\mathbf{x_1},\ldots,\mathbf{x_N}\} \in \mathbb{R}^d$, labels $\{y_1, \ldots, y_N\}$, loss function $\mathcal{L}(\theta)=\frac{1}{N}\sum_i \mathcal{L}(f_{\theta}(\mathbf{x_i}), y_i)$. Parameters: learning rate $\eta_t$, noise scale $\sigma$, group size $L$, gradient norm bound $C$. 
%\bmnote{Give a formula for choosing $\sigma_t$. Also: why two Input: (\REQUIRE) lines?}
%LZ: we are adopting the strategy of choosing sigma and then using privacy
%accountant to accumulate the privacy loss. Maybe we should make that more
%explicit in Privacy Accountant section?
		\STATE {\bf Initialize} $\theta_0$ randomly
		\FOR{$t \in [T]$}
		\STATE {Take a random sample $L_t$ of examples with sampling probability $q = L/N$}
		\STATE {\bf Compute gradient}
		\STATE {For each $i\in L_t$, compute $\mathbf{g}_t(\mathbf{x_i}) \gets \nabla_{\theta_t} \mathcal{L}(f_{\theta_t}(\mathbf{x_i}), y_i)$}	
		\STATE {\bf Clip gradient}
		\STATE {$\bar{\mathbf{g}}_t(\mathbf{x_i}) \gets \mathbf{g}_t(\mathbf{x_i}) / \max\big(1, \frac{\|\mathbf{g}_t(\mathbf{x_i})\|_2}{C}\big)$}
		\STATE {\bf Add noise}
		\STATE {$\tilde{\mathbf{g}}_t \gets \frac{1}{L}\left( \sum_i \bar{\mathbf{g}}_t(\mathbf{x_i}) + \mathcal{N}(0, \sigma^2 C^2 \mathbf{I})\right)$}
		\STATE {\bf Descent}
		\STATE { $\theta_{t+1} \gets \theta_{t} - \eta_t \tilde{\mathbf{g}}_t$}
		\ENDFOR
		\STATE {\bf Output} $\theta_T$ and compute the overall privacy cost $(\varepsilon, \delta)$ using a privacy accounting method.
%\bmnote{Ref to Eqs or pseudocode for computing privacy cost.}
%LZ: we will spend most of time discussing how privacy is computed. Since
%the algorithm description is an outline, maybe we can postpone it?
	\end{algorithmic}
\end{algorithm}

\textbf{Differential Privacy Under Post-Processing} \quad If $M$ is $\varepsilon$-DP, and $G$ is an arbitrary deterministic mapping, then $G \circ M$ is also $\varepsilon$-DP. In other words, once a quantity is ``made private" through DP, it cannot be subsequently ``made un-private" \cite{dwork}.

It is by post-processing of DP that an $(\varepsilon,\delta)$-DP fine-tuned model gives rise to $(\varepsilon,\delta)$-DP post-hoc explanations on the fine-tuning dataset, since explanations of the form $\varphi(\mathcal{X}, f_{\theta}, \mathbf{x}, \cdot)$ are functions of the model $f_{\theta}$. In this section, we present the existing DP optimiziations we use that improve upon the utility and computational efficiency of Abadi et al.'s DP-SGD algorithm.

\textbf{DP Utility and Efficiency Optimizations: Automatic Gradient Norm Clipping} \quad Algorithm \ref{alg:dpsgd}, which shows the DP-SGD algorithm, contains a gradient L2 norm clipping step in order to bound the global sensitivity of each batch gradient to a fixed, pre-determined parameter $C$: 
$\bar{\mathbf{g}}_t(\mathbf{x_i}) \gets \mathbf{g}_t(\mathbf{x_i}) \cdot \min\big(1, \frac{C}{\|\mathbf{g}_t(\mathbf{x_i})\|_2}\big)$. There are two main downsides of this gradient norm clipping method:
\begin{enumerate}
    \item We must tune $C$ (treated as a hyperparameter). This adds overhead to computational cost, especially since differentially private deep learning already also involves tuning parameters such as $\varepsilon$, $\delta$, batch size, sampling rate, and learning rate.
    \item The vanilla clipping method makes gradients lose important information. 
    Consider the per-sample gradient equivalent of Abadi et al.'s batch gradient clipping: $\bar{\mathbf{g}}(\mathbf{x_i}) \leftarrow \mathbf{g}(\mathbf{x_i}) \cdot \min\big(1, \frac{C}{\|\mathbf{g}(\mathbf{x_i})\|_2}\big)$. All per-sample gradients that have norm above $C$ are ``punished equally" and end up with the same magnitude: $|| \mathbf{g}(\mathbf{x_i}) \cdot \min\big(1, \frac{C}{\|\mathbf{g}(\mathbf{x_i})\|_2}\big)|| = C, \forall \mathbf{x_i}$.
\end{enumerate}

In this work, we instead employ \citet{bu2023automatic} \textit{automatic} gradient norm clipping algorithm to fine-tune models to improve model accuracy. The automatic per-sample gradient clipping works as follows:
\begin{equation}
    \bar{\mathbf{g}}_t(\mathbf{x_i}) \gets \mathbf{g}_t(\mathbf{x_i}) \cdot \frac{1}{\|\mathbf{g}_t(\mathbf{x_i})\|_2 + \gamma}, \label{equation:dpsgd_auto}
\end{equation}
where parameter $\gamma > 0$ is known as the ``stability constant." This approach remedies the two aforementioned downsides of vanilla gradient norm clipping. This approach allows gradients with larger norms to still have larger norms after clipping, whereas clipping will diminish the magnitudes of gradients with smaller norms. Furthermore, this clipping method frees us of the parameter $C$. Bu et al. prove that clipping with multiplicative factor $\displaystyle\frac{C}{\|\mathbf{g}_t(\mathbf{x_i})\|_2 + \gamma}$ is functionally equivalent to choosing $C=1$ (as in Equation \ref{equation:dpsgd_auto}), in the sense that in both cases, the gradient norm converges to zero at the same asymptotic rate.

We use Bu et al.'s Automatic Clipping to fine-tune vision transformers with DP.

\section{L1-LRT/L2-LRT Intuition and Algorithm}\label{appendix:l1_l2_lrt}

Before we discuss intuition on constructing likelihood ratio test statistics based on explanation norms, we first discuss intuition behind using gradient norms, since gradients are closely related to explanations.

\noindent \textbf{Intuition: Attacks on Gradient Norms} \quad 
Broadly speaking, a model $f_{\theta}$ is trained to approximately minimize the loss that $f_{\theta}$ incurs on training examples. The gradient of the model loss with respect to model parameters reflects the magnitude and direction of the ``step" that gradient descent takes during model training. The following intuition assumes a convex loss function. As the training process approaches a local minimum of the ``loss landscape" (i.e. the structure of the loss function in the parameter space that the model traverses step-wise during training), the model takes smaller and smaller steps in each subsequent iteration of the gradient descent process, until it reaches convergence. A trained model is not as ``well-fit" to non-members of the training set as it is to members. Hence, the model is more likely to take steeper and bigger gradient descent steps on unseen test set examples than on train set examples. The \textit{L1 norm} of the gradient directly encodes the steepness of the descent step taken after a model ``sees" an example. This intuition suggests that the gradient norms of training set members are on average smaller than gradient norms of non-members. Although the cross-entropy loss function we use is not convex in complex neural network settings, this intuition still motivates us to experiment with the gradient norm attack method.

\noindent \textbf{Intuition: From Gradient Norms to Explanation Norms} \quad 
The backpropagation-based post-hoc explanation methods that we work with involve computing gradients of $f_{\theta}$'s model output with respect to input features. These gradients are not exactly the same as the gradients computed during training, which are gradients of the loss function with respect to model parameters. However, we can still leverage the aforementioned intuition, since model parameter values directly reflect---albeit in a non-linear manner---how input features contribute to model predictions. Furthermore, there is separate intuition on the type of gradient computed in post-hoc explanations: this gradient represents the extent to which $f_{\theta}$'s \textit{prediction} changes if we were to perturb the input features. Since the training process pushes training set members further away from the decision boundary compared to non-members (behavior we previously explained in Section \ref{section:methods}), it follows that perturbing the input features of an arbitrary \textit{training} data point would scarcely change the model's behavior on or prediction for that point. Conceptually, this corresponds to a smaller gradient magnitude on training points---and magnitudes are equivalent to \textit{L2 norms}. 

\textbf{The Attack Algorithm} \quad Algorithm \ref{alg:norm_lrt} shows our explanation L1 norm-based LiRA method (L1-LRT). The L2-LRT attack is almost identical but instead based instead on L2 norms.

\begin{algorithm}[htb!]
 \begin{algorithmic}[1] 
  \REQUIRE \text{model} $f_{\theta}$, \text{example} $(\mathbf{x}, y) \in \mathbb{R}^d$, \text{explanation vector} $\varphi(f, (\mathbf{x}, y)) \in \mathbb{R}^d$, \text{data distribution} $\mathbb{D}$, \text{number of shadow model iterations} $N_S$
  \STATE $\text{norms}_{\text{in}} = \{\}$
  \STATE $\text{norms}_{\text{out}} = \{\}$
  \FOR{$N_S$ times}
    \STATE $D_{\text{attack}} \gets^\$ \mathbb{D}$ \algcomment{sample a shadow dataset}
    \STATE $f_{\text{in}} \gets \mathcal{T}(D_{\text{attack}} \cup \{(\mathbf{x},y)\})$ \algcomment{train IN model with $(\mathbf{x}, y)$ in training set}
    \STATE $\varphi_{\text{in}} \gets \varphi(f_{\text{in}}, (\mathbf{x}, y))$
    \algcomment{generate post-hoc explanation of $f_{\text{in}}$'s behavior on $(\mathbf{x}, y)$}
    % \STATE $\bar{\varphi}_{\text{in}} \gets \frac{1}{d} \sum_{i=1}^{d} \varphi_{\text{in}, i}$
    \STATE $\text{norms}_{\text{in}} \gets \text{norms}_{\text{in}} \cup \{ \sum_{i=1}^{d} |\varphi_{\text{in}, i}| \}$
    \algcomment{record L1 norm of $\varphi_{\text{in}}$}
    \STATE $f_{\text{out}} \gets \mathcal{T}(D_{\text{attack}} {\setminus \{(\mathbf{x},y)\}})$ \algcomment{train OUT model with $(\mathbf{x}, y)$ not in training set}
    \STATE $\varphi_{\text{out}} \gets \varphi(f_{\text{out}}, (\mathbf{x}, y))$
    \algcomment{generate post-hoc explanation of $f_{\text{out}}$'s behavior on $(\mathbf{x}, y)$}
    % \STATE $\bar{\varphi}_{\text{out}} \gets \frac{1}{d} \sum_{i=1}^{d} \varphi_{\text{out}, i}$
    \STATE $\text{norms}_{\text{out}} \gets \text{norms}_{\text{out}} \cup \{ \sum_{i=1}^{d} |\varphi_{\text{out}, i}| \}$
     \algcomment{record L1 norm of $\varphi_{\text{out}}$}
  \ENDFOR
  \STATE $\hat{\mu}_{\text{in}} \gets \texttt{mean}(\text{norms}_{\text{in}})$
  %{1 \over N}\sum_{i=1}^N \text{loss}_{\text{in}}$
  \STATE $\hat{\mu}_{\text{out}} \gets \texttt{mean}(\text{norms}_{\text{out}})$
  %{1 \over N}\sum_{i=1}^N \text{loss}_{\text{in}}$
  \STATE $\hat{\sigma}_{\text{in}}^2 \gets \texttt{var}(\text{norms}_{\text{in}})$
  %{1 \over N}\sum_{i=1}^N \text{loss}_{\text{in}}$
  \STATE $\hat{\sigma}_{\text{out}}^2 \gets \texttt{var}(\text{norms}_{\text{out}})$
  %{1 \over N}\sum_{i=1}^N \text{loss}_{\text{in}}$
  \STATE $\varphi_{\text{obs}} \gets \varphi(f_{\theta}, (\mathbf{x}, y))$
  \STATE $\text{norm}_{\text{obs}} = \sum_{i=1}^{d} |\varphi_{\text{obs}, i}| $ \algcomment{query target model}
  \vspace{0.5em}
  \STATE \textbf{return} $\displaystyle \hat{\Lambda} = \frac{p(\text{norm}_{\text{obs}}\ \mid\ \mathcal{N}(\hat{\mu}_{\text{in}}, \hat{\sigma}^2_{\text{in}}))}
    { p(\text{norm}_{\text{obs}}\ \mid\ \mathcal{N}(\hat{\mu}_{\text{out}}, \hat{\sigma}^2_{\text{out}}))}$
 \end{algorithmic}
 \caption{\textbf{L1-LRT: Likelihood ratio attack on the L1 norm of post-hoc explanations.}
 The adversary trains shadow models on datasets with and without the target example, generates post-hoc explanations on each example in their dataset, estimates parameters of the in- and out- distributions of sample variances of post-hoc explanations, and runs
 a likelihood ratio test.
 }
 \label{alg:norm_lrt}
\end{algorithm}

\section{Experimental Setups and Implementation Details}\label{appendix:experimental_setup}

\subsection{Datasets}
In this section, we discuss experimental setups and implementation details.

Across models and datasets, we sub-sample a smaller dataset of size $20000$ for fine-tuning each shadow model and computing post-hoc explanations. In the membership inference attack literature, sub-sampling is commonplace. In our predecessor work, \citet{featureattack} employ sub-sampling in many of their experiments on explanation-based membership inference attacks, using sub-sample sizes of $5000$, $10000$, and $20000$, among others. We generally use a $50\%/50\%$ train-test split across all attack and shadow model training procedures, since our attack success evaluation metrics are most straightforward to interpret when there is a balanced amount of training and test data given to the adversary; this is also the approach taken by Shokri et al. 

We present results for models fine-tuned on the following datasets designed for image classification. Each dataset consists of color images in $3$ color channels (red, green, and blue).

\textbf{CIFAR-10 and CIFAR-100} \quad CIFAR-$10$ and CIFAR-$100$ \cite{cifar} are well-known and widely used benchmark datasets for image classification. They consist of $10$ and $100$ classes, respectively, with $6000$ and $600$ images per class, respectively. The datasets are by default split into $50000$ training images and $10000$ test images, but for purposes of our membership inference attacks, we use a $50\%/50\%$ train-test split.

\textbf{Food 101} \quad Food 101 \cite{food101} is a dataset of 101 food categories with 101,000 images in total. For \textit{each class}, there are 750 training and 250 test images. According to Boassard et al., ``on purpose, the training images were not cleaned, and thus still contain some amount of noise. This comes mostly in the form of intense colors and sometimes wrong labels."

\textbf{Street View House Numbers (SVHN)} \quad The SVHN dataset \cite{svhn_data} contains satellite images of house numbers in Google Street View. It is similar to MNIST \cite{mnist} in that images are of small cropped digits and that there are $10$ classes, but it is a larger dataset ($73257$ train and $26032$ test images) and contains color images (whereas MNIST images are black-and-white). SVHN's increased complexity (compared to MNIST) makes it an appropriate downstream task for pre-trained foundation models.

\textbf{German Traffic Sign Recognition Benchmark (GTSRB)} \quad The GTSRB dataset \cite{gtsrb_data} features $43$ classes of traffic signs split into $39209$ training images and $12630$ test images.

Each of our datasets is housed in Torchvision's \href{https://pytorch.org/vision/main/datasets.html}{datasets} module \cite{torchvision} (see \href{https://pytorch.org/vision/stable/generated/torchvision.datasets.CIFAR10.html}{CIFAR-10}, \href{https://pytorch.org/vision/main/generated/torchvision.datasets.CIFAR100.html}{CIFAR-100}, \href{https://pytorch.org/vision/main/generated/torchvision.datasets.Food101.html}{Food 101}, \href{https://pytorch.org/vision/main/generated/torchvision.datasets.SVHN.html}{SVHN}, \href{https://pytorch.org/vision/stable/generated/torchvision.datasets.GTSRB.html}{GTSRB}).

\subsection{Model Architectures and Training}
We import and fine-tune pre-trained models from \href{https://pytorch.org/vision/stable/generated/torchvision.datasets.GTSRB.html}{\texttt{timm}} (standing for Py\textbf{T}orch \textbf{I}mage \textbf{M}odels) \cite{timm}, a deep learning library that provides state-of-the-art computer vision models and helper utilities to work with them. 

For each dataset, we experiment across the following model architectures. Each model has a patch size of 16, an input image dimension of 224, and is pre-trained on some ordered sample (possibly with replacement) of ImageNet-22k, ImageNet-21k, and ImageNet-1k.
\begin{itemize}
    \item CIFAR-$10$: \texttt{timm}'s \href{https://huggingface.co/timm/vit_small_patch16_224.augreg_in21k}{\texttt{vit\_small\_patch16\_224}} (30.1 M parameters), \href{https://huggingface.co/timm/vit_relpos_small_patch16_224.sw_in1k}{\texttt{vit\_relpos\_small\_patch16\_224.sw\_in1k}} (22.0 M parameters, with relative position embeddings), and \href{https://huggingface.co/timm/vit_relpos_base_patch16_224.sw_in1k}{\texttt{vit\_relpos\_base\_patch16\_224.sw\_in1k}} (86.4 M parameters, with relative position embeddings). In the main body, we report all attacks on the \texttt{vit\_small\_patch16\_224} model.
    \item CIFAR-$100$: \texttt{timm}'s \href{https://huggingface.co/timm/beit_base_patch16_224.in22k_ft_in22k_in1k}{\texttt{beit\_base\_patch16\_224.in22k\_ft\_in22k\_in1k}} (86.5 M parameters) and \\ \href{https://huggingface.co/timm/beitv2_base_patch16_224.in1k_ft_in22k_in1k}{\texttt{beitv2\_base\_patch16\_224.in1k\_ft\_in22k\_in1k}} (86.5 M parameters). For the main body, we report all attacks on \texttt{beit\_base\_patch16\_224.in22k\_ft\_in22k\_in1k}.
    \item Food 101, SVHN, and GTSRB: \href{https://huggingface.co/timm/vit_small_patch16_224.augreg_in21k}{\texttt{vit\_small\_patch16\_224}} and  \href{https://huggingface.co/timm/vit_relpos_small_patch16_224.sw_in1k}{\texttt{vit\_relpos\_small\_patch16\_224.sw\_in1k}}. For the main body, we report all attacks on \texttt{vit\_small\_patch16\_224}.
\end{itemize}

\subsection{Data Pre-Processing}\label{appendix:data_preprocessing}
We employ the following pre-processing methods for each image in each dataset:
\begin{enumerate}
    \item We resize each input image to have dimension $3 \times 224 \times 224$, where the first dimension corresponds to the three color channels (Red, Green, Blue). The per-color channel dimension is $224$ because that is the input dimension expected from the model architectures we use.
    \item We apply the transformation \texttt{torchvision.transforms.Normalize((0.5, 0.5, 0.5),(0.5, 0.5, 0.5))}. The first \texttt{(0.5, 0.5, 0.5)} corresponds to post-normalization mean of pixel values for each of the three (RGB) color channels, and the second \texttt{(0.5, 0.5, 0.5)} corresponds to post-normalization standard deviation.  This operation centers the input image tensors around zero and scales them to a range of approximately $-1$ to $1$.
\end{enumerate}

\subsection{Training Hyperparameters}\label{appendix:training_hyperparameters}

Table \ref{tab:training_hyperparameters} describes the chosen hyperparameter settings for each dataset, based on a combination of test accuracy (for model usefulness) and MIA attack success. ``Mini-batch size" describes the number of samples in each mini-batch during training; the model is trained on each mini-batch separately. ``Batch size" determines the sampling rate used in gradient descent. Sampling rate = (batch size) / (length of training data), and this quantity describes the proportion of the training data used for each parameter update step. This sampling rate is relevant to DP-SGD, where Gaussian noise is added to the gradients computed from only a subset of the training data at each update step.

\begin{table}[htb!]
\centering
\caption{Training hyperparameters for each dataset.}
\resizebox{0.6\columnwidth}{!}{
 \begin{tabular}{r|cccc}
    Dataset  & Batch Size & 
     Mini-Batch Size & Learning Rate & Epochs \\\hline
  CIFAR-$10$ & $1000$ & $50$ & $0.005$ & $30$\\
  CIFAR-$100$ & $1000$ & $50$ & $0.005$ & $9$\\
  Food 101 & $512$ & $50$ & $0.005$ & $50$\\
  SVHN & $512$ & $50$ & $0.005$ & $50$\\
  GTSRB & $512$ & $50$ & $0.005$ & $50$ \\
 \end{tabular}
 }
 \label{tab:training_hyperparameters}
\end{table}

\subsection{Post-Hoc Explanation Parameters}\label{appendix:explanation_parameters}

We use \href{https://github.com/pytorch/captum?tab=readme-ov-file}{Captum}, a model interpretability and understanding library for PyTorch \cite{captum}, to compute explanations in the form of attribution vectors. Captum supports all of the backpropagation-based methods we study (IXG, SL, IG, and GS), among others. Throughout this work, if an ROC curve or table omits results of a few particular settings of dataset and explanation type, that means it takes our computing resources too long to generate explanations of that type of $20000$ data examples.

In Captum, each feature attribution method accepts a list of parameters. Each method requires as input the \texttt{target} parameter, which specifies the output indices for which we want gradients to be computed. Captum's documentation \cite{captum} states that ``for classification cases, this is usually the target class." We retain this default (\texttt{target} = predicted class), with the intuition that the explanations should capture the features important to the model's predictions on the predicted class, not on any other class.

Integrated gradients (IG) has a \texttt{baseline} parameter (see Section \ref{section:integrated_gradients}). We set this $\mathbf{x}_{BL}$ quantity to the all-zero tensor, which is the default value in the Captum library. IG also has an \texttt{n\_steps} parameter, which describes the number of approximation steps used in integration. Captum sets the default \texttt{n\_steps} value to $50$, but to speed up computation, we set \texttt{n\_steps} $ = 25$.

Gradient SHAP (GS) has a \texttt{baseline} parameter as well (see Section \ref{section:gradient_shap}), which we set to a tensor where each component is distributed $\mathcal{N}(0, 0.001^2)$. GS also has an \texttt{n\_samples} parameter used for the following, according to Captum's documentation: ``[GS] adds white noise to each input sample \texttt{n\_samples} times, selects a random baseline from baselines’ distribution and a random point along the path between the baseline and the input, and computes the gradient of outputs with respect to those selected random points." Captum sets \texttt{n\_samples} to $5$ by default, and we retain this setting.

\subsection{Likelihood Ratio Attack Implementation}\label{section:lira_implementation}
For each attack setting, we train $N+1$ total models, where $N$ is the total number of shadow models of each attack. We perform $N+1$ runs of each attack, each time treating a different model as the target model and treating the remaining $N$ models as shadow models. Each of the $N+1$ models is trained on a randomly selected $10000$ points out of the subsampled dataset of size $20000$, and the remaining $10000$ points are used for testing. For each of the $20000$ examples, we record whether that example is in the training set or the test set of each model and save that information as a vector of $0$'s and $1$'s. The \textit{training set membership information} of all $N+1$ models is saved in a matrix of dimension $20000 \times (N+1)$. For each example, we also record the variance, L1 norm, and L2 norm ``scores" of each model's post-hoc explanation of that example. The \textit{explanation scores} of all examples on all $N+1$ models are saved in three matrices, each of dimension $20000 \times (N+1)$. Using the saved explanation scores (variances and L1/L2 norms) and training set membership statuses of each example, we run likelihood ratio attacks. 

For the attacks on models fine-tuned \textit{without} privacy, use $N=32$, meaning that we train $32$ shadow models per attack and perform $33$ total runs of each attack setting. For the attacks  on models fine-tuned \textit{with} privacy, we use $N=16$, meaning that we train $16$ shadow models per attack and perform $17$ total runs of each attack setting. We set $N$ to be lower here to respect fair and timely use of computational resources, considering the high amount of time it takes to generate post-hoc explanations and the multiple privacy settings (i.e. values of $\varepsilon$) we experiment on.

\subsection{Differential Privacy Parameters and Fine-Tuning} 

For all attacks on explanations coming from models fine-tuned with DP, we report results across the following values of privacy parameter $\varepsilon$: 0.5, 1.0, 2.0, 8.0, $\infty$. For conciseness, we report results involving DP on only the \texttt{vit\_small\_patch16\_224} and \texttt{beit\_base\_patch16\_224.in22k\_ft\_in22k\_in1k} models and the CIFAR-10 and CIFAR-100 datasets. In addition, we found the GS explanation method and the GTSRB dataset to yield low baseline (non-private) attack success; in order to provide more interpretable and distinctive results, we thus exclude these settings from results shown in this work. With privacy, ROC curves and metric values are averaged over 17 evaluation runs rather than 33, and each attack uses 16 shadow models rather than 32. This holds too for results on non-private model-based attacks in this section, in order to do an apples-to-apples comparison of non-private and private fine-tuning. As a result, in results involving DP, the reported baseline (non-private model) ROC curves and metric values differ by an extremely small amount from those reported in results not involving DP. Nonetheless, the results preserve the success of baseline attacks as well as the relative success of baseline attacks across datasets and explanation types. 

\section{Performance of Non-Private Models} \label{appendix:model_performance}
In Table \ref{tab:accuracies}, we present train and test accuracies for non-privately fine-tuned models on all datasets. We observe that different models fine-tuned on the same datasets have similar test accuracies on these datasets.

 \begin{table}[h!]
  \centering
 \caption[Model train and test accuracies.]{Model performance. We report average train and test accuracies for all non-privately finetuned models on all datasets. The ``chosen" epoch counts are shown as bolded rows. The results are averaged over 33 evaluation runs and include $\pm 1$ standard deviation.}
 \resizebox{0.75\columnwidth}{!}{
 \begin{tabular}{r|ccc}
    $\text{CIFAR-10}$  & Train Accuracy (\%) & Test Accuracy (\%) \\\hline
   \texttt{vit\_small\_patch26\_224} & $100.000 \pm 0.000$ & $96.064 \pm  0.613$ \\
   \texttt{vit\_relpos\_small\_patch16\_224.sw\_in1k} & $99.938 \pm 0.0797$ & $95.404 \pm  0.691$ \\
   \texttt{vit\_relpos\_base\_patch16\_224.sw\_in1k} & $99.788 \pm 0.202$ & $95.508 \pm  0.831$ \\
 \end{tabular}
 }
 
\resizebox{0.75\columnwidth}{!}{
 \begin{tabular}{r|ccc}
    $\text{CIFAR-100}$  & Train Accuracy (\%) & Test Accuracy (\%) \\\hline
  \texttt{beit\_base\_patch16\_224.in22k\_ft\_in22k\_in1k}& $98.722 \pm 0.321$ &  $80.109 \pm 0.590$ \\
\texttt{beitv2\_base\_patch16\_224.in12k\_ft\_in22k\_in1k}& $94.928 \pm 0.748$ &  $81.902 \pm 0.621$ \\
  
 \end{tabular}
}

\resizebox{0.75\columnwidth}{!}{
\begin{tabular}{r|ccc}
    $\text{Food 101}$  & Train Accuracy (\%) & Test Accuracy (\%) \\\hline
   \texttt{vit\_small\_patch26\_224} & $99.761 \pm 0.325$ & $83.685 \pm  3.878$ \\
   \texttt{vit\_relpos\_small\_patch16\_224.sw\_in1k} & $99.630 \pm 0.429$ & $81.204 \pm  3.139$ \\
 \end{tabular}
 }

\resizebox{0.75\columnwidth}{!}{
 \begin{tabular}{r|ccc}
    $\text{SVHN}$  & Train Accuracy (\%) & Test Accuracy (\%) \\\hline
   \texttt{vit\_small\_patch26\_224} & $99.552 \pm 0.223$ & $91.558 \pm  1.056$ \\
   \texttt{vit\_relpos\_small\_patch16\_224.sw\_in1k} & $99.580 \pm 0.215$ & $91.823 \pm  1.177$ \\
 \end{tabular}
 }

\resizebox{0.75\columnwidth}{!}{
 \begin{tabular}{r|ccc}
    $\text{GTSRB}$  & Train Accuracy (\%) & Test Accuracy (\%) \\\hline
   \texttt{vit\_small\_patch26\_224} & $100.000 \pm 0.000$ & $99.899 \pm  0.037$ \\
   \texttt{vit\_relpos\_small\_patch16\_224.sw\_in1k} & $100.000 \pm 0.000$ & $99.912 \pm  0.030$ \\
 \end{tabular}
 }
 \label{tab:accuracies}
\end{table}

\section{More Non-Private VAR-LRT Results}\label{appendix:additional_var_lrt}

\subsection{Comparing VAR-LRT with Thresholding Attack: More Table \ref{tab:lira_vs_simple_variance} Analysis}\label{appendix:green_red_table_explanation}

With respect to Table \ref{tab:lira_vs_simple_variance}, we observe that a majority of the red $\Delta$ values correspond to the AUC metric. This is the metric with which we are least concerned, for the following reasons: 
\begin{enumerate}
    \item We established in Section \ref{section:preliminaries} that AUC is an average-case metric that we inherently care about less than we care about TPR at low FPR.
    \item All of the red $\Delta_{\texttt{AUC}}$ values correspond to Gradient SHAP (GS) attacks. With GS attacks, we observe that both attack types on GS generally have low success: in the GS rows, we see average $\texttt{TPR}_{.001}$ values closest to $0.001$ and average $\texttt{TPR}_{.01}$ values closest to $0.01$ (compared to attacks on other explanation types). The conclusion that both attacks are minimally successful is more salient than any conclusion made comparing the two attacks. 
    % \item The red $\Delta_{\texttt{AUC}}$ values appear most often in the GTSRB dataset. Relatively, both attacks perform worse on GTSRB than on our other datasets, and objectively, both attacks are unsuccessful on GTSRB. Hence, we weigh GTSRB results with relatively lesser importance compared to results on other datasets.
\end{enumerate}

For these reasons, we are not concerned about negative $\Delta_{\texttt{AUC}}$ values, so we use a darker shade of red to color them. The singular other entry with a brighter shade of red indicates that the CIFAR-$10$ VAR-LRT attack based on Gradient SHAP (GS) performs weaker than the corresponding thresholding attack in the $\texttt{TPR}_{.001}$ metric. However, we claim that this entry \textit{also} does not undermine the relative success of VAR-LRT: the bright red $\Delta_{\texttt{TPR}_{.001}}$ value shows a difference that is not statistically significant. The $t$-statistic of the two-sided hypothesis test comparing the mean $\Delta_{\texttt{TPR}_{.001}}$ values of the two attacks is $-1.946$, and the corresponding $p$-value is $0.0561$, which is considered not low, assuming a $0.05$-level test.

Almost all of the green-colored $\Delta$ values show statistically significant differences. In particular, the green $\Delta$ values corresponding to the CIFAR-$10$, SVHN, and CIFAR-$100$ datasets and the IXG, SL, and IG explanation methods are all significant. (The test statistics and $p$-values of the two-sample $t$-tests verifying this conclusion are in Table \ref{tab:p_values_varlrt_vs_thresholding}.) These are the very datasets and explanation methods we care about most, since both attack methods are more successful than random guessing in these settings.

\begin{table}[htb!]
\centering
\caption[Two-sample $t$-test results comparing VAR-LRT and thresholding attack performance.]{Two-sample $t$-test results comparing VAR-LRT and thresholding attack performance.
We report $t$ statistics and $p$-values for two-sample, two-sided $t$-tests of each metric under the CIFAR-$10$, CIFAR-$100$, Food 101, and SVHN datasets and the IXG, SL, and IG explanation types. In each setting, the mean difference in metric value between the VAR-LRT and thresholding attacks is positive, meaning VAR-LRT exhibits a statistically significant performance improvement compared to the thresholding attack. The sample size is $33$ for each hypothesis test.}
\resizebox{0.85\columnwidth}{!}{ %
\begin{tabular}{clcccccccc}
\toprule
\multirow{2}{*}{Explanation Type} & \multirow{2}{*}{Metric} & \multicolumn{2}{c}{CIFAR-$10$} &  \multicolumn{2}{c}{SVHN} & \multicolumn{2}{c}{CIFAR-$100$} & \multicolumn{2}{c}{Food 101}\\
\cmidrule(l){3-4} \cmidrule(l){5-6} \cmidrule(l){7-8} \cmidrule(l){9-10}
& & $t$-statistic & $p$-value & $t$-statistic & $p$-value & $t$-statistic & $p$-value & $t$-statistic & $p$-value \\
\cmidrule(lr){1-1} \cmidrule(lr){2-2} \cmidrule(l){3-4} \cmidrule(l){5-6} \cmidrule(l){7-8} \cmidrule(l){9-10}
\multirow{3}{*}{IXG} 
& $\text{TPR}_{.001}$ & 
$16.291$ & $1.876\mathrm{E}{-24}$ & $6.092$ & $7.050\mathrm{E}{-08}$ & $9.166$ & $2.909\mathrm{E}{-13}$ &  $7.300$ & $5.533\mathrm{E}{-10}$ \\
& $\text{TPR}_{.01}$ & 
$40.594$ & $2.146\mathrm{E}{-47}$ & $11.183$ & $1.072\mathrm{E}{-16}$ & $ 22.176$ & $ 9.645 \mathrm{E}{-32}$ & $11.734$ & $1.334\mathrm{E}{-17}$ \\
& AUC & 
$29.013$ & $1.534\mathrm{E}{-38}$ &  $12.007$ & $4.815\mathrm{E}{-18}$ & $ 5.934$ & $1.312\mathrm{E}{-07}$ & $5.082$ & $3.486\mathrm{E}{-6}$\\
\cmidrule(lr){1-10}
\multirow{3}{*}{SL}
 & $\text{TPR}_{.001}$ & 
 $13.216$ & $5.914\mathrm{E}{-20}$ &  $6.341$ & $2.621\mathrm{E}{-08}$ & $ 9.977 $ & $ 1.156\mathrm{E}{-14}$ & $7.185$ & $8.787\mathrm{E}{-10}$ \\
& $\text{TPR}_{.01}$ & 
$30.706$ & $5.176\mathrm{E}{-40}$ &  $12.507$ & $7.639\mathrm{E}{-19}$ & $22.674 $ & $2.725\mathrm{E}{-32}$ & $11.345$ & $5.796\mathrm{E}{-17}$\\
 & AUC & 
 $24.003$ & $1.039\mathrm{E}{-33}$ & $12.561$ & $6.257\mathrm{E}{-19}$ & $ 6.029$ & $ 9.032 \mathrm{E}{-08}$ & $4.900$ & $6.857\mathrm{E}{-6}$ \\
\cmidrule(lr){1-10}
\multirow{3}{*}{IG} 
& $\text{TPR}_{.001}$ & 
$7.233$ & $7.248\mathrm{E}{-10}$ &  $7.049$ & $1.526\mathrm{E}{-09}$ & - & - & $9.156$ & $ 3.027 \mathrm{E}{-13}$ \\
& $\text{TPR}_{.01}$ &
$25.241$ & $5.625\mathrm{E}{-35}$ & $11.745$ & $1.279\mathrm{E}{-17}$ & - & -  & $ 12.105$ & $ 3.346\mathrm{E}{-18}$\\
& AUC & 
$16.740$ & $4.566\mathrm{E}{-25}$ & $9.931$ & $1.382\mathrm{E}{-14}$ & - & - & $ 4.861$ & $7.910\mathrm{E}{-6}$ \\
\bottomrule
\end{tabular} 
}

\label{tab:p_values_varlrt_vs_thresholding}
\end{table}

Hence, the red values in Table \ref{tab:lira_vs_simple_variance} do not undermine the conclusion that VAR-LRT is a stronger attack than the thresholding attack, particularly at confidently identifying specific members of the training dataset. Further, this result holds across datasets and explanation methods.

\subsection{Results on More Models and Datasets}

\begin{figure}[htb!]
    \centering
\includegraphics[width=0.24\linewidth]{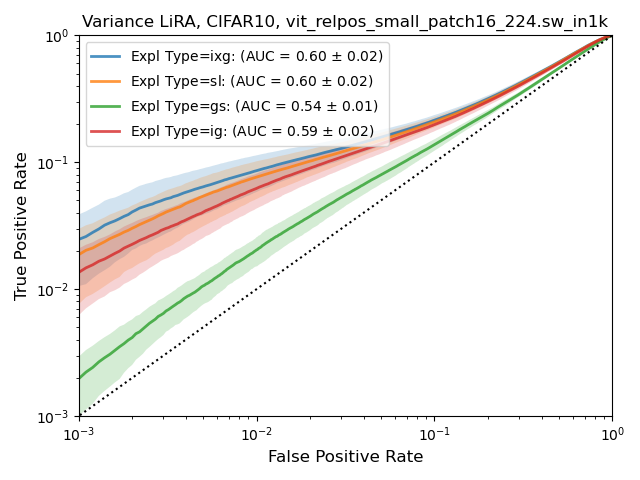}\includegraphics[width=0.24\linewidth]{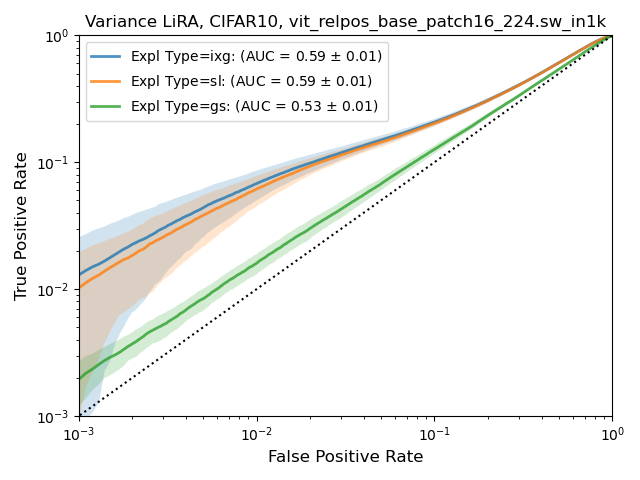}\includegraphics[width=0.24\linewidth]{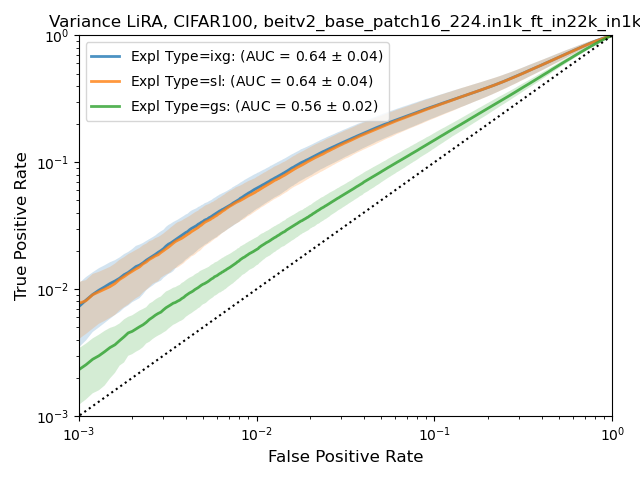}
\includegraphics[width=0.24\linewidth]{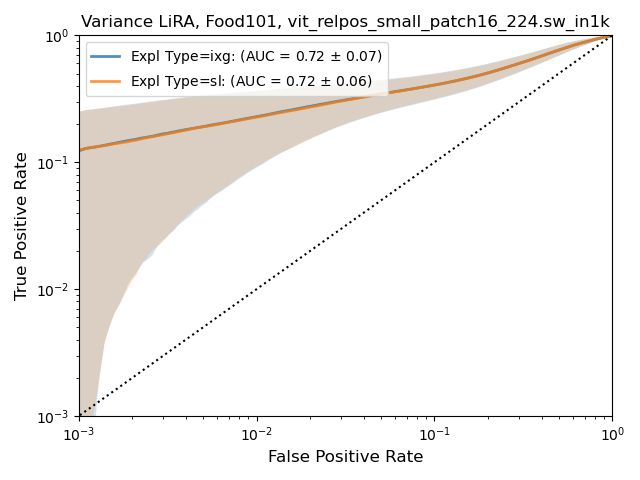}

    \caption{VAR-LRT log-scaled ROC curves for the CIFAR-10 (first and second from the left), CIFAR-100 (second from the right), and Food 101 (right) datasets, on different model architectures than are presented in the main body.}
    \label{fig:variance_nondp_other_models}
\end{figure}

Figure \ref{fig:variance_nondp_other_models} shows VAR-LRT ROCs for CIFAR-10, CIFAR-100, and Food 101 under the additional model architectures \textit{not} shown in the main body. We observe that the VAR-LRT attack remains successful on this new set of vision transformer models---particularly at low FPR.

\begin{itemize}
    \item CIFAR-$10$: \texttt{vit\_relpos\_small\_patch16\_224.sw\_in1k, vit\_relpos\_base\_patch16\_224.sw\_in1k}
    \item CIFAR-$100$: \texttt{beit\_base\_patch16\_224.in22k\_ft\_in22k\_in1k}
    \item Food 101: \texttt{vit\_relpos\_small\_patch16\_224.sw\_in1k}
\end{itemize}

\begin{figure}[htb!]
    % \begin{subfigure}{\linewidth}
    \centering
    \includegraphics[width=0.32\linewidth]{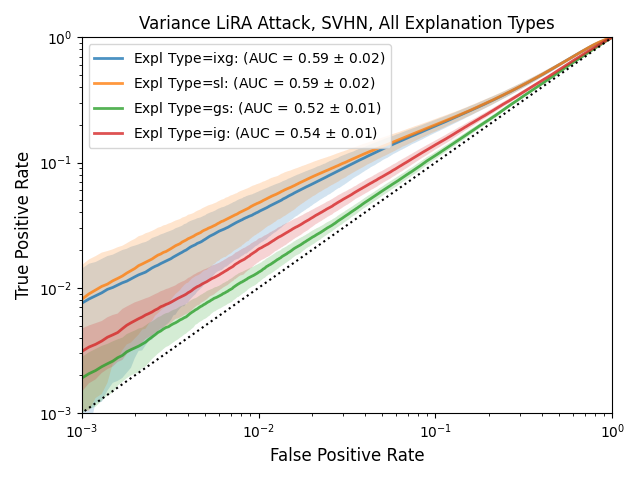} \includegraphics[width=0.32\linewidth]{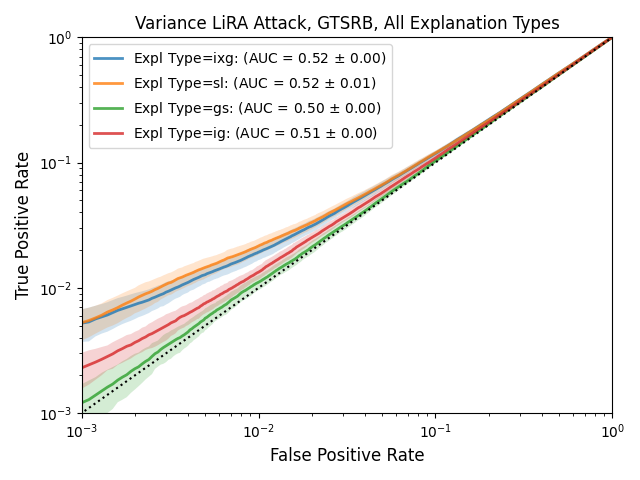}

    % \caption{SVHN (left) and GTSRB (right).}
    % \end{subfigure}
    
    \caption{VAR-LRT ROCs for the SVHN (left) and GTSRB (right) datasets, \texttt{vit\_small\_patch16\_224} model.}
    \label{fig:variance_nondp_svhn_gtsrb}
\end{figure}

In the main text, we presented VAR-LRT attack ROCs for the CIFAR-$10$, CIFAR-$100$, and Food 101 datasets but excluded plots on the SVHN and GTSRB datasets. Figure \ref{fig:variance_nondp_svhn_gtsrb} shows these excluded plots, using the \texttt{vit\_small\_patch16\_224} architecture. VAR-LRT also performs better than random guessing on these datasets, particularly at low FPR.

\section{More Non-Private L1-LRT/L2-LRT Results}\label{appendix:additional_l1_l2_lrt}

\begin{figure}[htb!]
    \centering

    \begin{subfigure}{\linewidth}
    \centering
\includegraphics[width=0.24\linewidth]{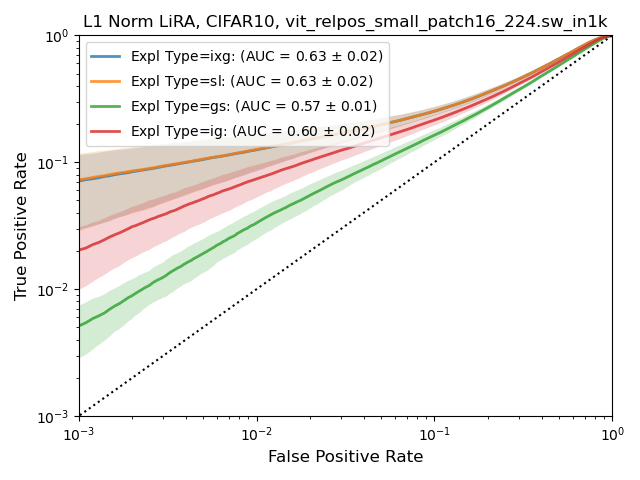}\includegraphics[width=0.24\linewidth]{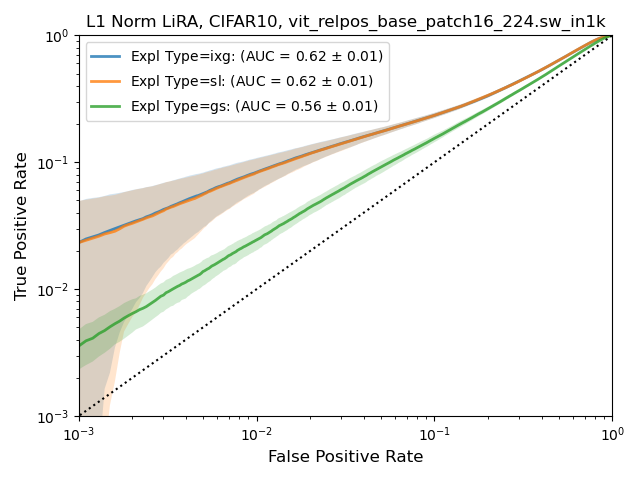}\includegraphics[width=0.24\linewidth]{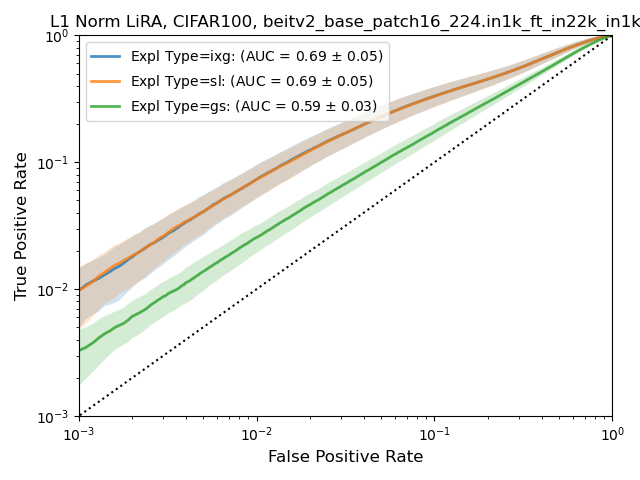}
\includegraphics[width=0.24\linewidth]{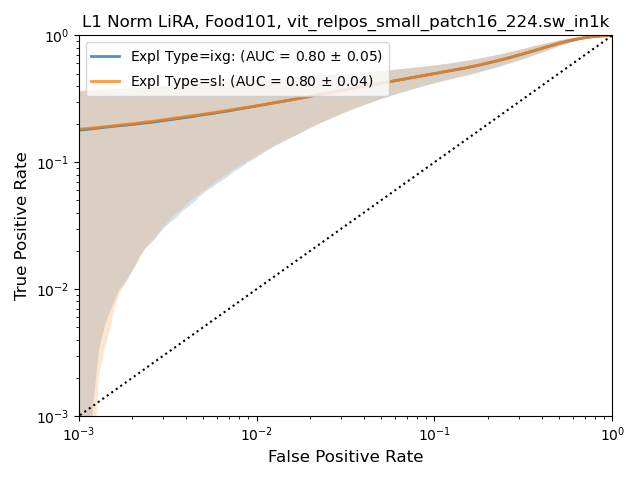}
\caption{L1-LRT log-scaled ROC curves.}
\end{subfigure}

    \begin{subfigure}{\linewidth}
    \centering
\includegraphics[width=0.24\linewidth]{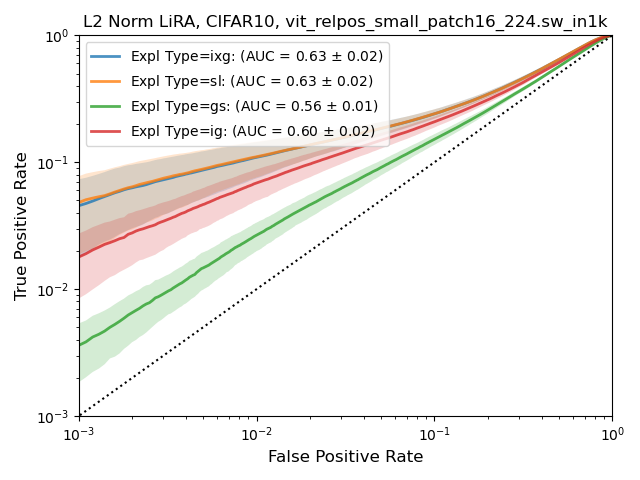}
\includegraphics[width=0.24\linewidth]{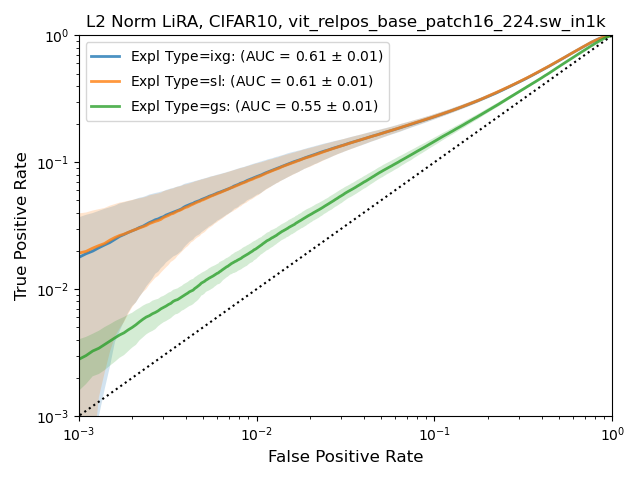}\includegraphics[width=0.24\linewidth]{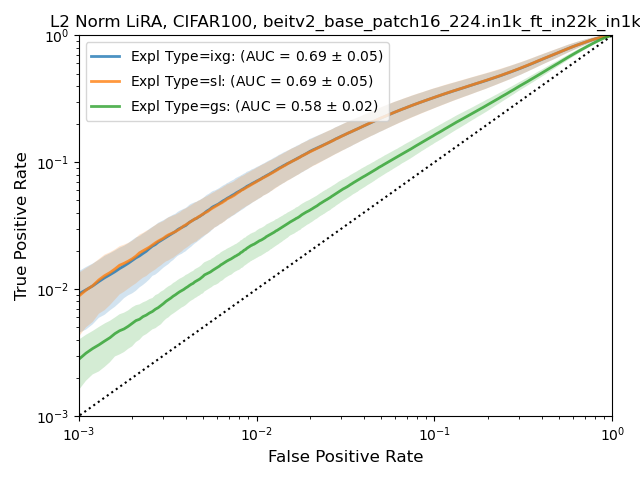}
\includegraphics[width=0.24\linewidth]{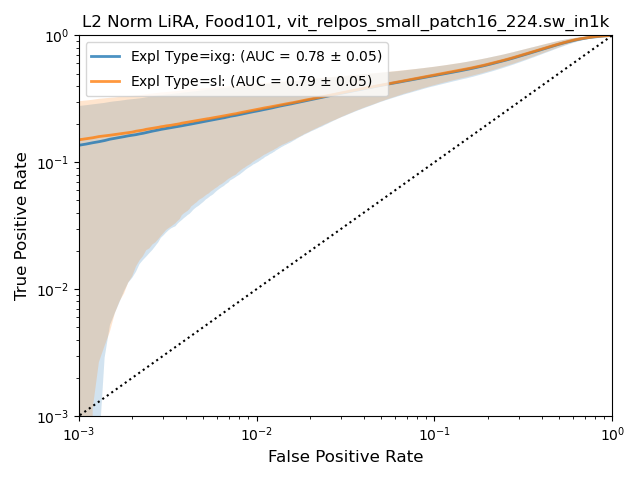}
\caption{L2-LRT log-scaled ROC curves.}
\end{subfigure}
    
    \caption{L1-LRT and L2-LRT attack results for the CIFAR-10 (first and second from the left), CIFAR-100 (second from the right), and Food 101 (right) datasets, on different model architectures than are presented in the main body.}
    \label{fig:l1_l2_norm_nondp_additional}
\end{figure}

In Figure \ref{fig:l1_l2_norm_nondp_additional}, we present L1-LRT (top) and L2-LRT (bottom) results on the following datasets and architectures that were not featured in the main body:

\begin{itemize}
    \item CIFAR-$10$: \texttt{vit\_relpos\_small\_patch16\_224.sw\_in1k, vit\_relpos\_base\_patch16\_224.sw\_in1k}
    \item CIFAR-$100$: \texttt{beit\_base\_patch16\_224.in22k\_ft\_in22k\_in1k}
    \item Food 101: \texttt{vit\_relpos\_small\_patch16\_224.sw\_in1k}
\end{itemize}

We observe that the L1-LRT and L2-LRT attacks remain successful on this new set of vision transformer models.

\begin{figure}[htb!]
    \centering
    \begin{subfigure}{\linewidth}
    \centering
    \includegraphics[width=0.32\linewidth]{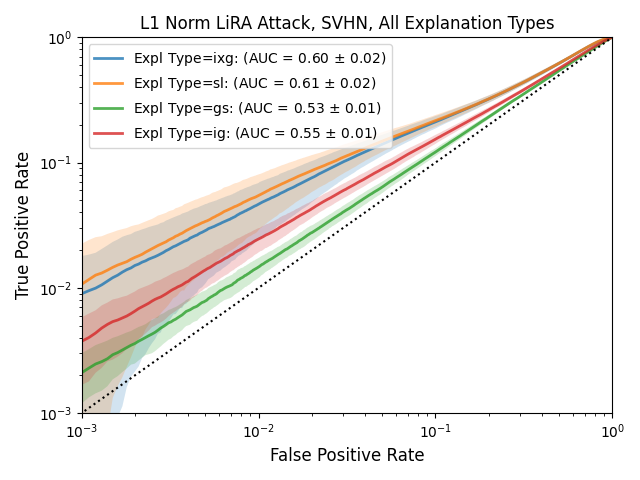} 
    \includegraphics[width=0.32\linewidth]{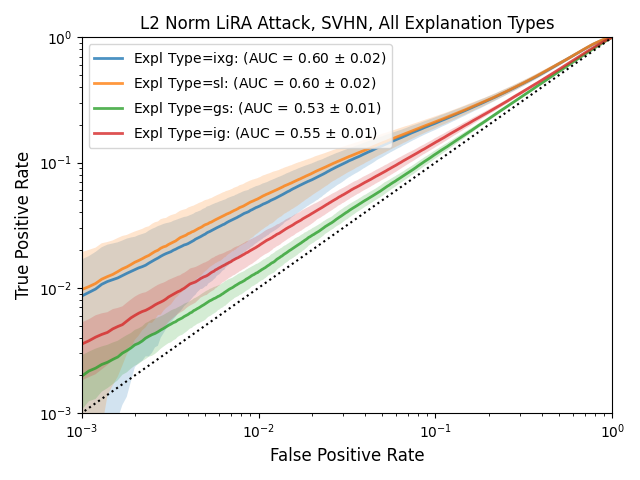}  
    \caption{SVHN, L1-LRT (left) and L2-LRT (right).}
    \end{subfigure}
    \begin{subfigure}{\linewidth}
    \centering
    \includegraphics[width=0.32\linewidth]{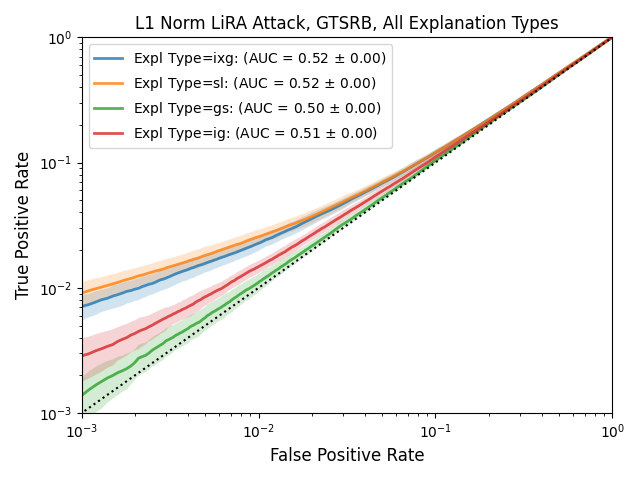}   \includegraphics[width=0.32\linewidth]{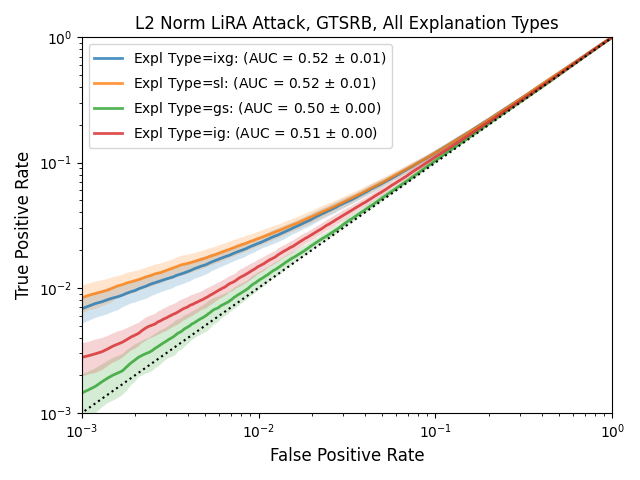} 

    \caption{GTSRB, L1-LRT (left) and L2-LRT (right).}
    \end{subfigure}
    \caption{L1-LRT and L2-LRT attack ROCs for the SVHN and GTSRB datasets, across all explanation types.}
    \label{fig:l1_l2_norm_nondp_svhn_gtsrb}
\end{figure}

In the main text, we presented L1-LRT and L2-LRT attack ROCs for the CIFAR-$10$, CIFAR-$100$, and Food 101 datasets but excluded plots on the SVHN and GTSRB datasets. In Figure \ref{fig:l1_l2_norm_nondp_svhn_gtsrb}, we show these excluded plots, coming from the \texttt{vit\_small\_patch16\_224} model. 

\subsection{Comparing L1-LRT and L2-LRT}\label{appendix:l1_lrt_vs_l2_lrt}
We observe in the main text, as well as in this appendix, that L1-LRT attacks are more successful than L2-LRT attacks overall. We hypothesize that this may be related to the gradient of the cross-entropy loss with respect to weights in the last hidden layer of the underlying model. Let $w_{ji}$ be the weight linking hidden unit value $h_j$ to the (pre-activation) output $z_i$: this means $z_i = h_j w_{ji} + b_j$, where $b_j$ is a bias term. For feature vector $\mathbf{x}$, let $y_i$ be the $i$th element of the ground-truth one-hot encoded vector $\textbf{y} \in \{0,1\}^k$. Let $\hat{y}_i = p(\mathbf{x})_i$ represent the $i$th element of the model's predicted probability distribution over the classes. \citet{beaujour} derives the gradient of cross-entropy loss with respect to weight $w_{ji}:$ $\displaystyle\frac{\partial \mathcal{L}}{\partial w_{ji}} = h_j(\hat{y}_i - y_i).$ The gradient is linear in the distance between predicted and true class probabilities, so intuitively, gradient descent ``travels linearly" through this probability vector space. 
The gradient of the model \textit{output} with respect to the input features is closely related to the gradient of the model \textit{loss} with respect to the final-layer weights, since model weights directly reflect how input features map to model predictions.

We thus hypothesize that the L1 norm of the gradient of the model output with respect to input features, which is also a ``linear" distance metric, better reflects the linear behavior of gradient descent than does the L2 norm of the gradient. However, this is but a hypothesis, and we encourage future exploration into this result.

\section{Non-Private Ablation Experiments}\label{appendix:ablation_experiments}

\subsection{On the Computational Efficiency versus the Privacy Risk of Explanations}
Our tables and figures present a salient observation that we have not yet verbalized: that across datasets, attacks on Input * Gradient (IXG) and Saliency (SL) generally perform best, while attacks on Integrated Gradients (IG) and Gradient SHAP (GS) tend to have lower success. This finding highlights a trade-off between the computational efficiency of an explanation method and its susceptibility to privacy attack; according to Table \ref{tab:explanation_times}, IXG and SL attributions are much faster to compute than GS and IG explanations, at least with respect to our compute resources and platform.

An attacker can more readily leverage explanation methods that are computationally efficient: our attacks require computing a full set of feature attributions based on each shadow model, and this process is significantly easier if we use more efficient explanation methods. Thus, this trade-off is itself a sign of privacy risk.

\begin{table}[htb!]
\centering
\caption[Comparing computational efficiency of explanation methods.]{Comparing computational efficiency of explanation methods. Time (mm:ss) taken for each explanation method to generate attributions for 200 CIFAR-$10$ examples.}
\resizebox{0.35\columnwidth}{!}{ %
\centering
    \begin{tabular}{ |c|c| } 
    \hline
    Explanation Type & Time ($200$ Iters) \\
    \hline
    \texttt{IXG} & $1$:$07$  \\
    \texttt{SL} &  $1$:$07$ \\
    \texttt{GS} & $6$:$24$  \\
    \texttt{IG} &  $13$:$12$ \\
    \hline
    \end{tabular}
  }

\label{tab:explanation_times}
\end{table}

On the thread of comparing explanation methods with one another, one redeeming observation, however, is that because GS and IG \textit{theoretically} satisfy desirable axiomatic properties that SL and IXG do not, the observation that GS and IG are less susceptible to privacy attack is auspicious from an axiomatic approach: the explanations with axiomatic properties are also better defended against privacy risk. (Recall that Appendix \ref{appendix:post_hoc_explanations} highlights these axiomatic properties.)

\subsection{On the Impact of Overfitting and Underfitting}

\begin{figure}[htb!]
    \centering

    \begin{subfigure}{\linewidth}
    \centering
    \includegraphics[width=0.3\linewidth]{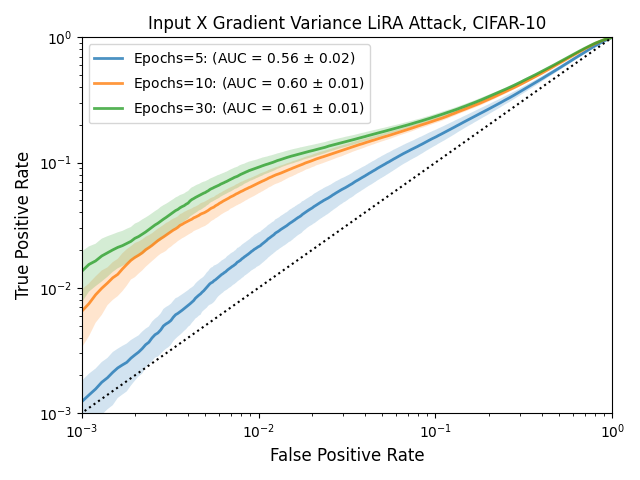} \includegraphics[width=0.3\linewidth]{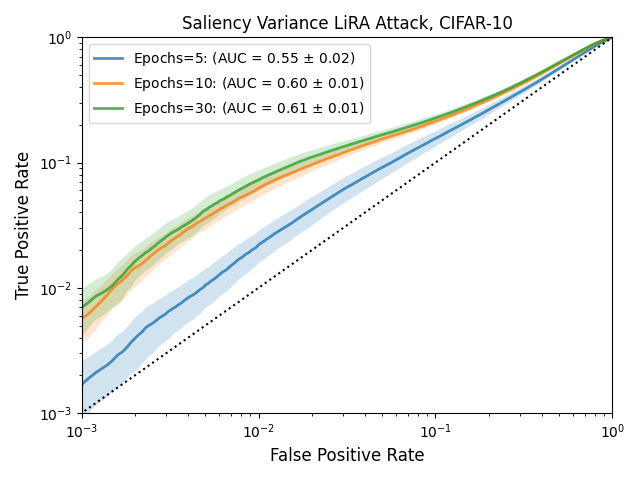} \includegraphics[width=0.3\linewidth]{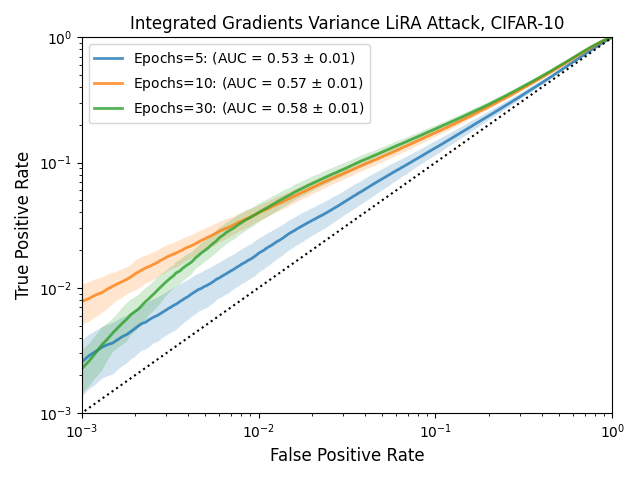} 

    \caption{CIFAR-10 Attacks.}
    \end{subfigure}
    \vfill

    \begin{subfigure}{\linewidth}
    \centering
    \includegraphics[width=0.3\linewidth]{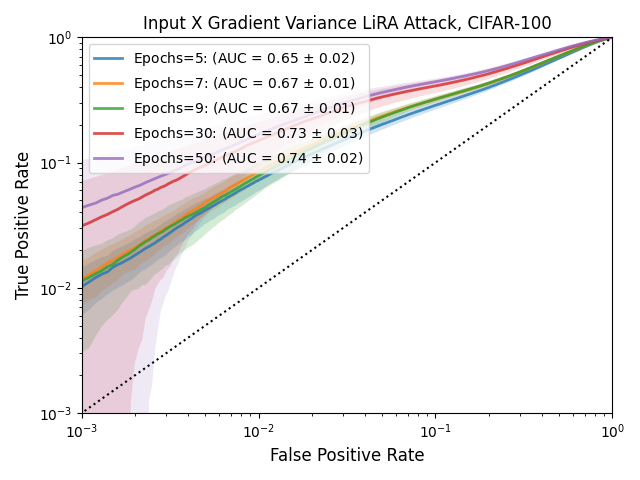} \includegraphics[width=0.3\linewidth]{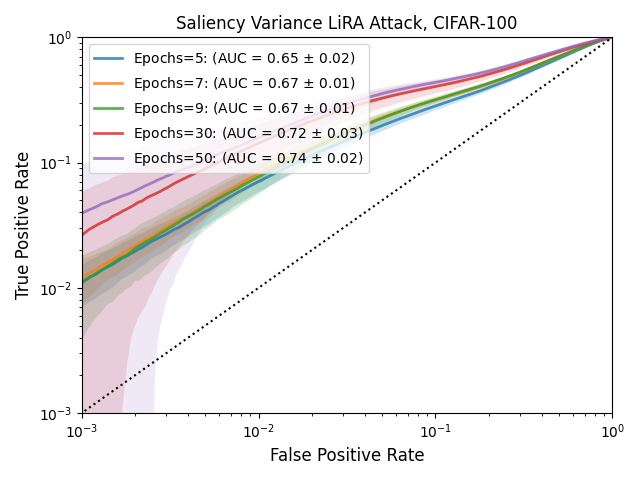}
    \caption{CIFAR-100 Attacks.}
    \end{subfigure}
    
    \caption{VAR-LRT on CIFAR-10 and CIFAR-100 data; multiple epoch settings. Each plot shows ROC curves of attacks for a single dataset and explanation type, with each curve within each plot corresponding to a different epoch setting.}
    \label{fig:multiple_epoch_settings}
\end{figure}

Figure \ref{fig:multiple_epoch_settings} presents attack performance plots across different fine-tuning epoch counts on CIFAR-10 and CIFAR-100 data using the \texttt{vit\_small\_patch16\_224} model.
We experiment across the following epoch counts for the two datasets:
\begin{itemize}
    \item CIFAR-10: 5, 10, 30 (30 is default)
    \item CIFAR-100: 5, 7, 9, 30, 50 (9 is default)
\end{itemize}

\textbf{CIFAR-10 and Underfitting} \quad Figure \ref{fig:multiple_epoch_settings} shows that even when the model is fine-tuned on CIFAR-$10$ for $10$ epochs (well below the ``chosen" $30$ epoch setting), VAR-LRT still performs successfully, at least compared with the thresholding attack, not only on average (through improved AUC) but especially in the FPR=$0.001$ and FPR=$0.01$ regions. Although we cannot make such strong statements about statistical significance for the $5$ epochs setting, the ROC curves and reported AUC values still show higher success for the VAR-LRT attack compared to the thresolding attack.

\textbf{CIFAR-100 and Overfitting} \quad Figure \ref{fig:multiple_epoch_settings} shows that even when the model is fine-tuned on CIFAR-$100$ for $30$ or $50$ epochs (well above the ``chosen" $9$ epoch setting), the ROC curves and reported AUC values show higher success for the VAR-LRT attack. Specifically, VAR-LRT captures significantly higher TPR than the thresholding attack when FPR is between $0.01$ and $0.1$. VAR-LRT shows improvement on average in other metrics (AUC, TPR \@ FPR=$0.001$) as well, albeit without statistical significance. Hence, VAR-LRT's performance exceeds that of the thresholding attack and is objectively successful even when the model is overfit or underfit. 

\textbf{Overfitting and Training Data Leakage} \quad This figure also shows that the longer we fine-tune a model for (that is, the more epochs the model is trained for), the more susceptible to privacy attack the ensuing explanations are; this result holds across datasets and explanation methods. Intuitively, the more epochs the model is trained for, the more ``familiar" the model becomes on training points, and the further away the decision boundary moves from these points. Explanations, by design, capture model behavior, and model behavior varies more between training and non-training examples as it becomes more ``familiar" with training examples. Hence, it follows intuitively that training data explanations will indeed on average differ more from non-training data explanations. This result more broadly reveals a downside to model overfitting beyond the more commonly discussed implication that overfitting leads to low model generalizability on unseen data: overfitting leads to \textit{increased data privacy risk}, especially as we add transparency to models through explainability. Consequently, as researchers investigate privacy risk defenses in model training, it is important and promising to consider approaches that directly or indirectly avoid overfitting.

\subsection{On the Impact of More Shadow Models}

\begin{figure}[htb!]
    \centering
    \includegraphics[width=0.4\linewidth]{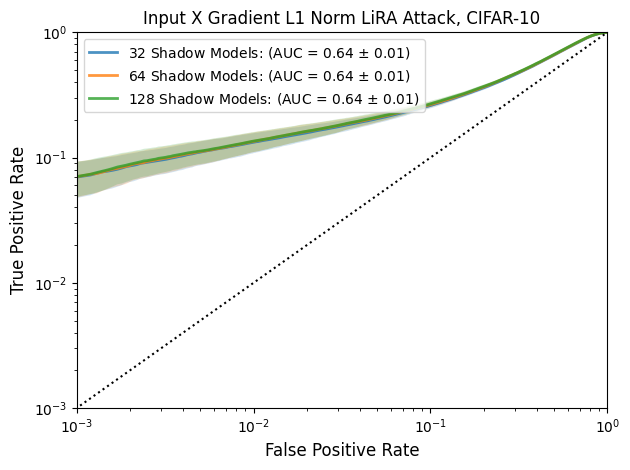} 
    \caption[Impact of changing the number of shadow models.]{Impact of changing the number of shadow models. We show log-scaled ROC curves for the IXG L1-LRT attack on CIFAR-10 over $[32, 64, 128]$ shadow models, using the \texttt{vit\_small\_patch16\_224} model. Each curve is taken across 20 evaluation runs. We observe no difference in attack performance after changing shadow model count.}\label{fig:changing_shadow_model_count}
\end{figure}

For this investigation, we fix the number of evaluation runs per attack setup to 20. Figure \ref{fig:changing_shadow_model_count} shows the IXG L1-LRT attack on CIFAR-10 and the \texttt{vit\_small\_patch16\_224} model over $[32, 64, 128]$ shadow models. We observe that even quadrupling the number of shadow models from 32 to 128 has no impact on attack performance. Several membership inference attack works evaluate their results with more shadow models than the 32 and 16 used in this work. For example, \citet{carlini} frequently use 64 and 128 shadow models in their experiments, and  \citet{abascal} use 128 shadow models. We use fewer shadow models out of respect for compute resource limitations, and Figure \ref{fig:changing_shadow_model_count} shows that we do not sacrifice on attack performance in doing so.

\section{Performance of Models Fine-Tuned with Differential Privacy} \label{appendix:dp_model_performance}

Table \ref{tab:dp_model_accuracies} shows average train and test accuracies for each dataset at DP $\varepsilon$ values of $0.5, 1.0, 2.0, 8.0$, and $\infty$ (non-DP). The table brings out a surprising observation: the test accuracies of some of the $\varepsilon$-DP models with higher $\varepsilon$ values are \textit{higher} than the test accuracy of the baseline non-private model. This is counter-intuitive and surprising: even at high values of $\varepsilon$, we inject noise into gradients during fine-tuning, so why would the model ever perform \textit{better} than if no random noise were injected into gradient computations? Formally, this should not happen. This observation is precisely why we report train accuracies \textit{alongside} test accuracies, even though we care about test accuracies more---train accuracies provide more perspective on this finding. The non-private models have considerably higher train accuracy than any of the private models but, perhaps as a result, they overfit more to the training data. The non-private models perform relatively poorly on training data, but it is \textit{by virtue} of the injected random noise that these models generalize better, yielding high test accuracy. The private models' performance relative to one another is expected according to the well-known privacy-utility tradeoff: the lower the value of $\varepsilon$ (i.e. the stronger the privacy protection), the lower the model accuracy.

Objectively, however, the DP fine-tuned models have high test accuracy and strong performance. We chose to run experiments on large pre-trained foundation models using automatic gradient norm clipping (in the DP algorithm) in the first place for this purpose: so that the privately fine-tuned models we work with have high utility.

\begin{table}[htb!]
\centering
\caption[Model train and test accuracies for each dataset at DP $\epsilon$ values of 0.5, 1.0, 2.0, 8.0, and $\infty$ (non-DP).]{Model train and test accuracies for each dataset at DP $\epsilon$ values of 0.5, 1.0, 2.0, 8.0, and $\infty$ (non-DP). Each model is fine-tuned with the chosen epoch setting highlighted in Table \ref{tab:accuracies}. We report the means computed over 17 independent runs, as well as the standard deviations.}
\resizebox{0.53\columnwidth}{!}{
\begin{tabular}{lccc}
\toprule [0.12em]
& \multirow{2}{*}{$\varepsilon$} & \multicolumn{2}{c}{Accuracy} \\
\cmidrule{3-4}
& & Train Accuracy (\%) & Test Accuracy (\%) \\
\midrule [0.075em]
\multirow{3}{*}{CIFAR-$10$} 
     & $0.5$ & $96.313 \pm 0.301$ & $95.796 \pm 0.148$ \\
     & $1.0$ & $97.123 \pm 0.196$ & $96.234 \pm 0.157$ \\
     & $2.0$ & $97.972 \pm 0.147$ & $96.496 \pm 0.162$ \\
     & $8.0$ & $99.021 \pm 0.101$ & $96.611 \pm 0.162$ \\
     & $\infty$ (non-DP) & $100.00 \pm 0.000$ & $96.064 \pm 0.613$ \\
\midrule
\multirow{3}{*}{CIFAR-$100$} 
     & $0.5$ & $73.754 \pm 1.262$ & $71.488 \pm 1.176$ \\
     & $1.0$ & $80.082 \pm 0.638$ & $77.346 \pm 0.767$ \\
     & $2.0$ & $83.305 \pm 0.566$ & $80.907 \pm 0.520$ \\
     & $8.0$ & $86.889 \pm 0.383$ & $84.343 \pm 0.321$ \\
     & $\infty$ (non-DP) & $98.722 \pm 0.321$ & $80.109 \pm 0.590$ \\
\midrule
\multirow{3}{*}{SVHN} 
     & $0.5$ & $76.708 \pm 2.066$ & $75.167 \pm 2.399$ \\
     & $1.0$ & $86.153 \pm 0.741$ & $84.149 \pm 0.989$ \\
     & $2.0$ & $89.694 \pm 0.378$ & $86.922 \pm 0.578$ \\
     & $8.0$ & $93.168 \pm 0.362$ & $88.922 \pm 0.406$ \\
     & $\infty$ (non-DP) & $99.542 \pm 0.223$ & $91.558 \pm 1.056$ \\
\bottomrule[0.12em]
\end{tabular}
}
\label{tab:dp_model_accuracies}
\end{table}

\section{More Results on the Impact of Differentially Private Fine-Tuning on Attack Success} \label{appendix:more_dp_results}

\subsection{VAR-LRT DP Results} 
\begin{figure}[htb!]
    \centering
    
    \begin{subfigure}{\linewidth}
    \centering
    \includegraphics[width=0.3\linewidth]{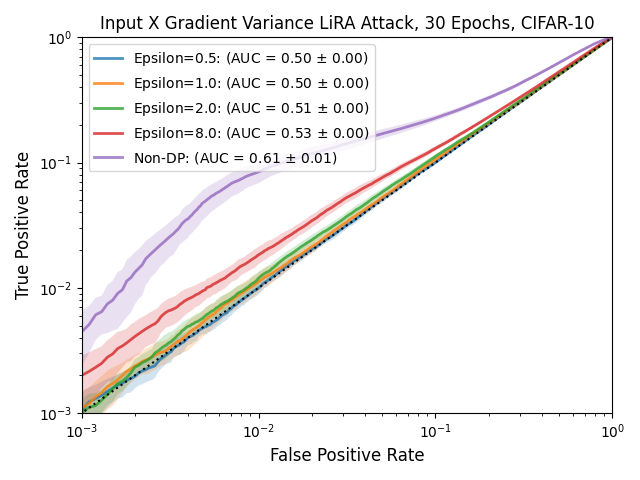} \includegraphics[width=0.3\linewidth]{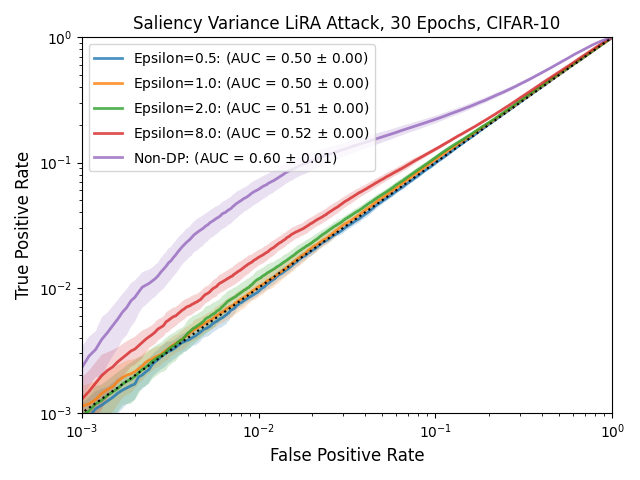} 
    \includegraphics[width=0.3\linewidth]{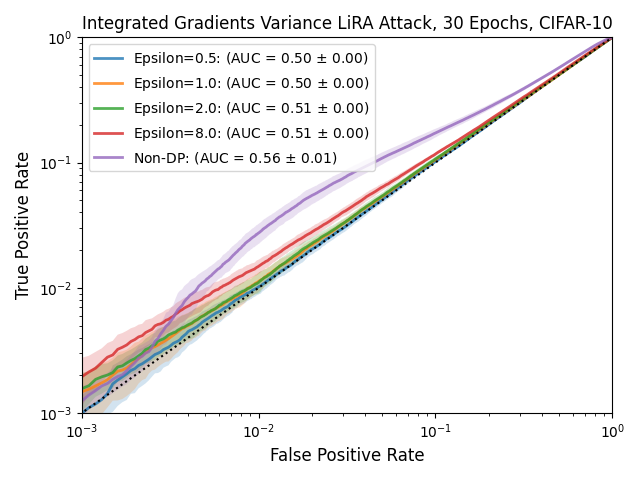} 
    \caption{CIFAR-10 (IXG, SL, IG).}
    \end{subfigure}
    \vfill

    \begin{subfigure}{\linewidth}
    \centering
    \includegraphics[width=0.3\linewidth]{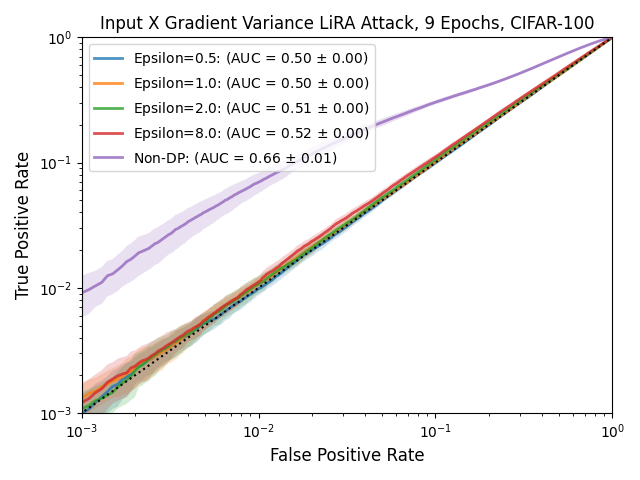} \includegraphics[width=0.3\linewidth]{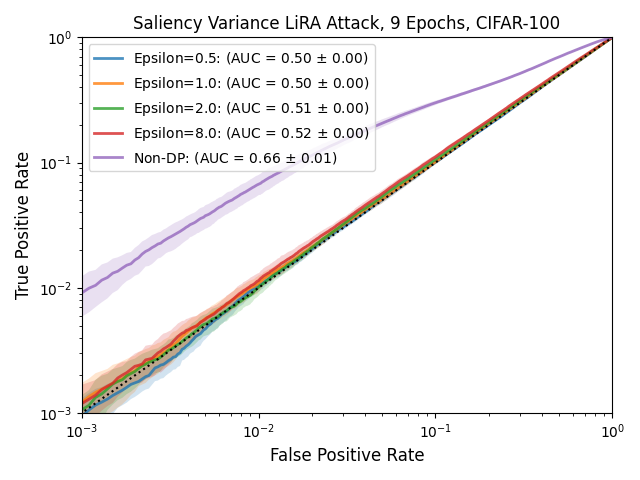} 
    \caption{CIFAR-100 (IXG, SL).}
    \end{subfigure}

    \vfill
    \begin{subfigure}{\linewidth}
    \centering
    \includegraphics[width=0.3\linewidth]{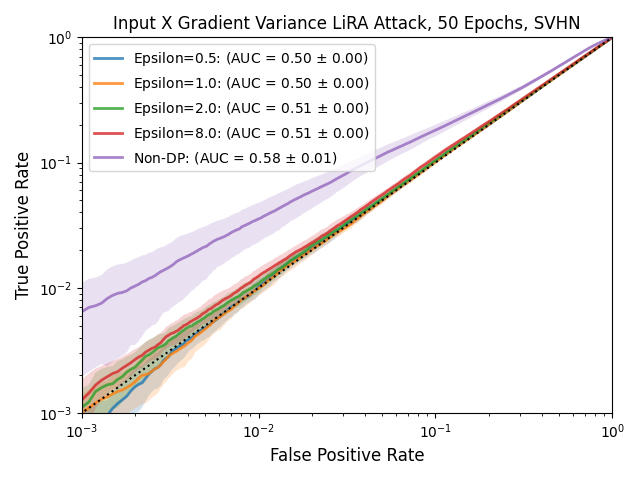} \includegraphics[width=0.3\linewidth]{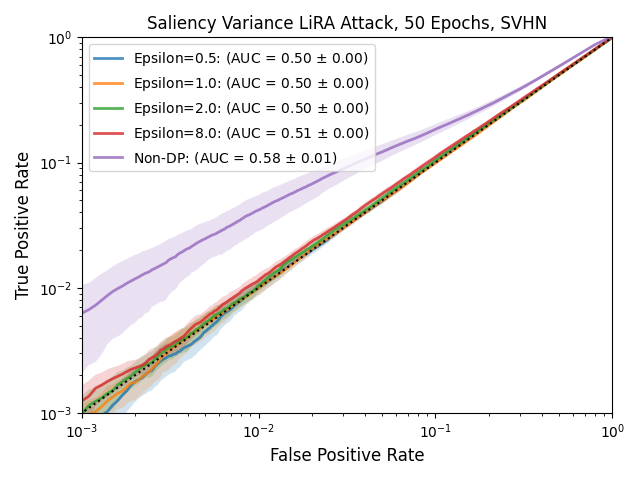} 
    \includegraphics[width=0.3\linewidth]{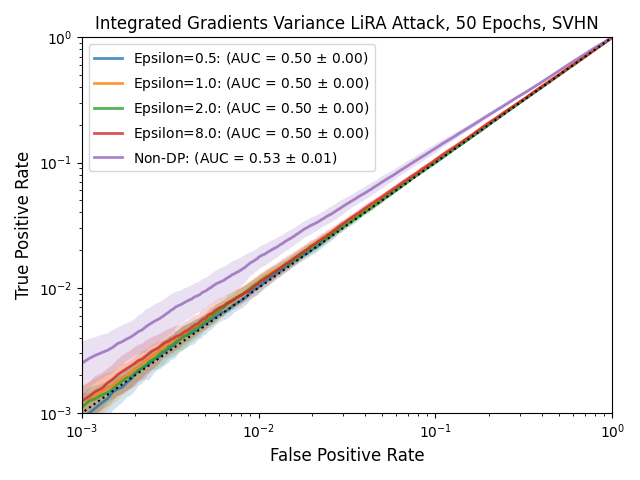} 
    \caption{SVHN (IXG, SL, IG).}
    \end{subfigure}
    
    \caption[VAR-LRT attack success of non-private vs. DP fine-tuned models.]{VAR-LRT attack success of non-private vs. DP fine-tuned models. We show one plot per explanation method: IXG (left), SL (middle), and (with the exception of CIFAR-100) IG (right). Each subplot shows curves for $\epsilon = 0.5, 1.0, 2.0, 8.0, \infty$.}
    \label{fig:variance_dp_all}
\end{figure}

Figure \ref{fig:variance_dp_all} shows log-scaled ROC curves highlighting the impact of DP fine-tuning on VAR-LRT attack success. Each subplot features one dataset and explanation type over different privacy settings. We observe that in each subplot---that is, across datasets and explanation types---the ROC curves corresponding to models fine-tuned with DP hug the ``random guessing" diagonal line much more closely than the baseline ROC curves do. For CIFAR-10, the $\varepsilon = 8.0$ setting shows minimal but nonzero attack success. However, even the least private $\varepsilon = 8.0$ setting yields unsuccessful attacks on CIFAR-100. Moreover, the other values of $\varepsilon=0.5, 1.0, 2.0$ (corresponding to stronger privacy protection settings) yield attacks that behave no better than random guessing, both on average and when FPR is low.

\subsection{L1/L2-LRT DP Results}
Table \ref{tab:best_attacks_dp_table} shows the $\text{TPR}_{.001}$, $\text{TPR}_{.01}$, and AUC L1-LRT attack metrics for the $\varepsilon=1.0, 2.0, 8.0$ privacy settings, the CIFAR-$10$ and CIFAR-$100$ datasets, and the IXG and SL explanation methods. For each metric, we also show results on non-private attacks $(\varepsilon=\infty$) over $16$ shadow models per run (and $17$ total runs). This table bolsters, from a numerical perspective, the conclusions we have drawn from the ROC curves in the main body.

\begin{table}[htb!]
\centering
\caption[L1-LRT attack success, DP vs. non-DP.]{L1-LRT attack success, DP vs. non-DP. 
We report the $\text{TPR}_{.001}$, $\text{TPR}_{.01}$, and AUC L1-LRT attack metrics for the $\varepsilon=1.0, 2.0, 8.0$ privacy settings, the CIFAR-10 and CIFAR-100 datasets, and the IXG and SL explanation methods. For each metric, we also show results on non-private attacks $(\varepsilon=\infty$). Results are averaged over 17 evaluation runs models per run (with 16 shadow models used per run).}
% \begin{subtable}{\linewidth}
%     \centering
\resizebox{0.8\columnwidth}{!}{ %
\begin{tabular}{cccccc}
\toprule
\multirow{2}{*}{Metric} & \multirow{2}{*}{Epsilon ($\varepsilon$)} & \multicolumn{2}{c}{CIFAR-$10$} &  \multicolumn{2}{c}{CIFAR-$100$} \\
\cmidrule(l){3-4} \cmidrule(l){5-6}
& & \texttt{IXG} & \texttt{SL} & \texttt{IXG} & \texttt{SL}  \\
\cmidrule(lr){1-1} \cmidrule(lr){2-2} \cmidrule(l){3-4} \cmidrule(l){5-6}
\multirow{3}{*}{$\text{TPR}_{.001} \ \downarrow$} 
& $1.0$ & 
$0.0019 \pm 0.0008$ & $0.0016 \pm 0.0007$ & $0.0013 \pm 0.0008$ & $0.0013 \pm 0.0007$ \\
& $2.0$ & 
$0.0019 \pm 0.0007$ & $0.0021 \pm 0.0011$ & $0.0012 \pm 0.0007$ & $0.0013 \pm 0.0009$ \\
& $8.0$ & 
$0.0040 \pm 0.0014$ & $0.0042 \pm 0.0016$ &  $0.0016 \pm 0.0008$ & $0.0015 \pm 0.0011$ \\
& $\infty$ (non-DP) & 
$0.0895 \pm 0.0213$ & $0.0902 \pm 0.0223$ &  $0.0225 \pm 0.0089$ & $0.0195 \pm 0.0079$ \\
\cmidrule(lr){1-6}
\multirow{3}{*}{$\text{TPR}_{.01} \ \downarrow$}
 & $1.0$ & 
 $0.0128 \pm 0.0023$ & $0.0126 \pm 0.0022$ &  $0.0109 \pm 0.0018$ & $0.0125 \pm 0.0022$ \\
& $2.0$ & 
$0.0159 \pm 0.0025$ & $0.0164 \pm 0.0029$ &  $0.0129 \pm 0.0025$ & $0.0115 \pm 0.0023$ \\
 & $8.0$ & 
 $0.0302 \pm 0.0044$ & $0.0305 \pm 0.0047$ & $0.0153 \pm 0.0031$ & $0.0154 \pm 0.0035$ \\
& $\infty$ (non-DP) & 
 $0.1530 \pm 0.0216$ & $0.1529 \pm 0.0230$ & $0.1332 \pm 0.0352$ & $0.1308 \pm 0.0033$ \\
\cmidrule(lr){1-6}
\multirow{3}{*}{\text{AUC} \ $\downarrow$} 
& $1.0$ & 
$0.5087 \pm 0.0035$ & $0.5092 \pm 0.0031$ &  $0.5018 \pm 0.0028$ & $0.5020 \pm 0.0032$ \\
& $2.0$ &
$0.5149 \pm 0.0040$ & $0.5142 \pm 0.0048$ & $0.5090 \pm 0.0048$ & $0.5100 \pm 0.0045$  \\
& $8.0$ & 
$0.5327 \pm 0.0044$ & $0.5325 \pm 0.0043$ & $0.5219 \pm 0.0041$ & $0.5226 \pm 0.0042$ \\
& $\infty$ (non-DP) & 
$0.6383 \pm 0.0095$ & $0.6383 \pm 0.0095$ & $0.7171 \pm 0.0084$ & $0.7153 \pm 0.0094$ \\
\bottomrule
\end{tabular} 
}
% \caption{CIFAR-10 (30 epochs) and SVHN (50 epochs).}
  %   \label{tab:subtable3}
  % \end{subtable}%

\label{tab:best_attacks_dp_table}
\end{table}

\begin{figure}[htb!]
    \centering
    \begin{subfigure}{\linewidth}
    \centering
    \includegraphics[width=0.3\linewidth]{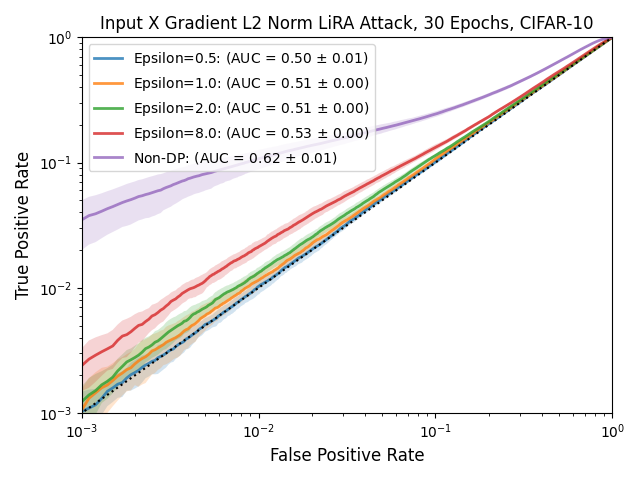}\includegraphics[width=0.3\linewidth]{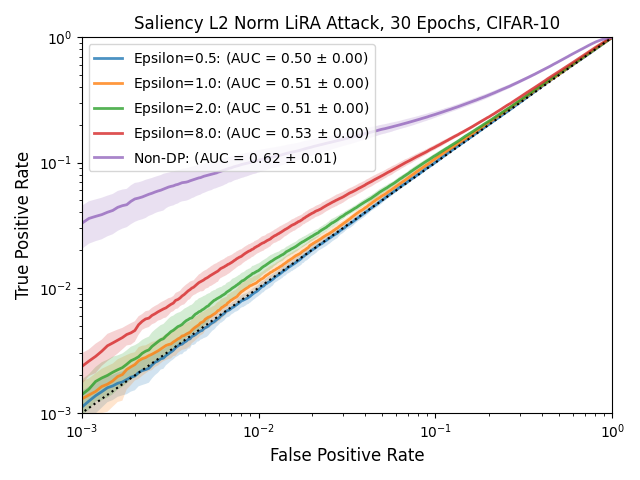}
\includegraphics[width=0.3\linewidth]{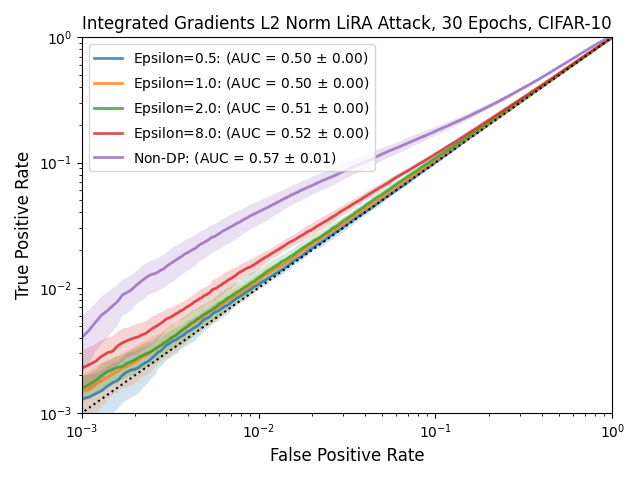}
    \caption{CIFAR-10 (IXG, SL, IG).}
    \end{subfigure}

    \vfill
    \begin{subfigure}{\linewidth}
    \centering
    \includegraphics[width=0.3\linewidth]{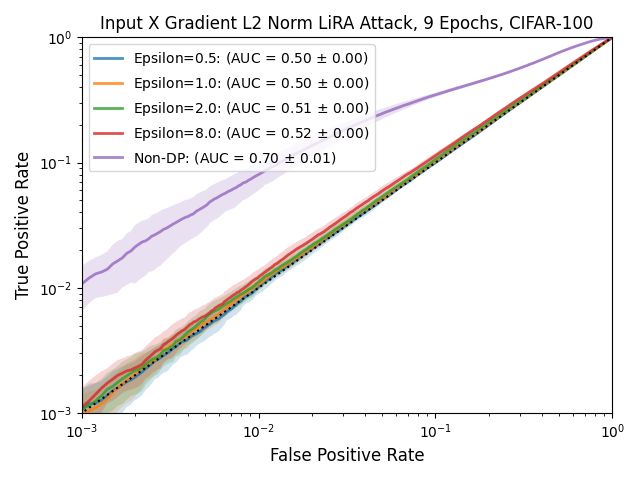}\includegraphics[width=0.3\linewidth]{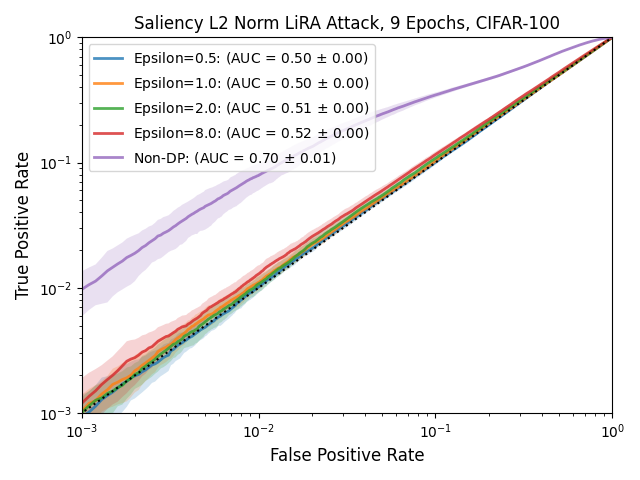}
    \caption{CIFAR-100 (IXG, SL).}
    \end{subfigure}

    \vfill
    \begin{subfigure}{\linewidth}
    \centering
   \includegraphics[width=0.3\linewidth]{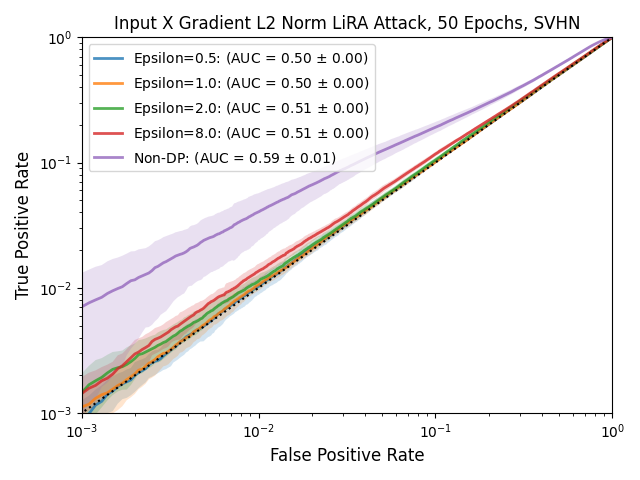}\includegraphics[width=0.3\linewidth]{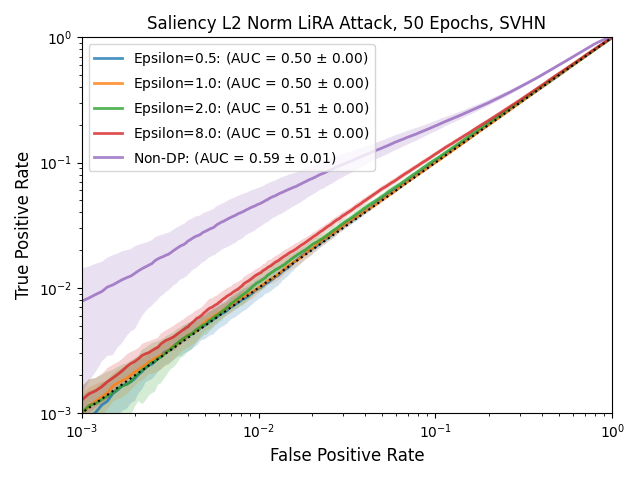}
\includegraphics[width=0.3\linewidth]{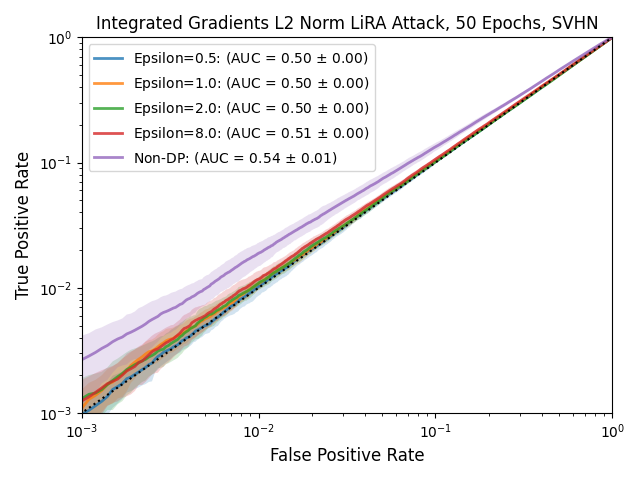}
    \caption{SVHN (IXG, SL, IG).}
    \end{subfigure}
    
    \caption[L2-LRT attack success of non-private vs. DP fine-tuned models.]{L2-LRT attack success of non-private vs. DP fine-tuned models. We show one plot per explanation method: IXG (left), SL (middle), and (with the exception of CIFAR-100) IG (right). Each subplot shows curves for $\epsilon = 0.5, 1.0, 2.0, 8.0$.}
    \label{fig:l2_norm_dp_all}
\end{figure}

Figure \ref{fig:l2_norm_dp_all} provides ROC curves showing L2-LRT attack success on non-private vs. DP fine-tuned models.

% You can have as much text here as you want. The main body must be at most $8$ pages long.
% For the final version, one more page can be added.
% If you want, you can use an appendix like this one.  

% The $\mathtt{\backslash onecolumn}$ command above can be kept in place if you prefer a one-column appendix, or can be removed if you prefer a two-column appendix.  Apart from this possible change, the style (font size, spacing, margins, page numbering, etc.) should be kept the same as the main body.
%%%%%%%%%%%%%%%%%%%%%%%%%%%%%%%%%%%%%%%%%%%%%%%%%%%%%%%%%%%%%%%%%%%%%%%%%%%%%%%
%%%%%%%%%%%%%%%%%%%%%%%%%%%%%%%%%%%%%%%%%%%%%%%%%%%%%%%%%%%%%%%%%%%%%%%%%%%%%%%

\end{document}